\documentclass[useAMs,usenatbib,a4paper]{mn2e}
\usepackage{savesym}
\usepackage{graphicx}
\expandafter\let\csname equation*\endcsname\relax
  \expandafter\let\csname endequation*\endcsname\relax 
\usepackage{subfig}
\usepackage{multirow}
\usepackage{amsmath}
\usepackage{verbatim}
\usepackage{array}
\usepackage{times}
\usepackage[total={17.8cm,24.0cm},centering]{geometry} 
\usepackage{rotating}
\usepackage{amssymb}
\usepackage{pdflscape}

\newcommand{\be}{\begin{equation}}
\newcommand{\ee}{\end{equation}}
\newcommand{\eea}{\end{eqnarray}}
\newcommand{\bea}{\begin{eqnarray}}

\newcommand{\m}{\mathrm}

\title[Understanding blazar flares with a fluid jet model]{Modelling blazar flaring using a time-dependent fluid jet emission model - an explanation for orphan flares and radio lags}
\author[William J. Potter]{William J. Potter\thanks{E-mail:
will.potter@astro.ox.ac.uk (WJP)}
\\
Oxford Astrophysics. Denys Wilkinson Building, Keble Road, Oxford, OX1 3RH, United Kingdom}
\begin{document}

\date{}

\pagerange{\pageref{firstpage}--\pageref{lastpage}} \pubyear{2011}

\maketitle

\label{firstpage}

\begin{abstract}
Blazar jets are renowned for their rapid violent variability and multiwavelength flares, however, the physical processes responsible for these flares are not well understood. In this paper we develop a time-dependent inhomogeneous fluid jet emission model for blazars.  We model optically thick radio flares for the first time and show that they are delayed with respect to the prompt optically thin emission by $\sim$ months to decades, with a lag that increases with the jet power and observed wavelength. This lag is caused by a combination of the travel time of the flaring plasma to the optically thin radio emitting sections of the jet and the slow rise time of the radio flare. We predict two types of flares: {\it symmetric flares} - with the same rise and decay time, which occur for flares whose duration is shorter than both the radiative lifetime and the geometric path-length delay timescale; {\it extended flares} - whose luminosity tracks the power of particle acceleration in the flare, which occur for flares with a duration longer than both the radiative lifetime and geometric delay. Our model naturally produces orphan X-ray and $\gamma$-ray flares. These are caused by flares which are only observable above the quiescent jet emission in a narrow band of frequencies. Our model is able to successfully fit to the observed multiwavelength flaring spectra and lightcurves of PKS1502+106 across all wavelengths, using a transient flaring front located within the broad-line region. 

\end{abstract}


\section{Introduction}

Blazars are the most luminous and highly variable sub-class of active galactic nuclei (AGN). Their relativistic plasma jets point close to Earth, resulting in substantial Doppler boosting of their emission \cite{1995PASP..107..803U}. This Doppler enhancement means that the radiation from blazars is dominated by jet emission making them invaluable for understanding jet physics and high energy particle acceleration. They are well-known for their rapid, powerful multiwavelength variability with the shortest flares unresolved on minute timescales and dramatically increasing in luminosity (\citealt{2007ApJ...664L..71A} and \citealt{2007ApJ...669..862A}). These flares have been the subject of numerous multiwavelength observational campaigns (\citealt{2005ApJ...630..130B}, \citealt{2011ApJ...729....2A}, \citealt{2011MNRAS.410..368B}, \citealt{2016ApJS..222....6B}, \citealt{2016ApJ...819..156B} etc.) because the spectral and temporal changes in emission properties are directly related to the particle acceleration mechanism operating in jets, which despite its central importance, is still not understood.

The observed flaring behaviour of blazars is difficult to model because of the erratic nature of the flares. The variability of some flaring events appears to be largely stochastic, with no clear link between variability at different wavelengths (e.g. \citealt{2011ApJ...738...25A} and \citealt{2012A&A...542A.100A}), whilst other flares have a distinct luminosity profile observable at multiple wavelengths, sometimes with frequency dependent lags (e.g. \citealt{2010ApJ...710..810A}). Previous work modelling flaring behaviour has focused on using one or two physically disconnected spherical emission regions to model flares, with varying success depending on the properties of the flare (e.g. \citealt{1998A&A...333..452K} \citealt{2000ApJ...536..729L} \citealt{2002ApJ...581..127B}, \citealt{2005ApJ...630..130B}, \citealt{2008ApJ...686..181F}, \citealt{2011A&A...534A..86T} \citealt{2012ApJ...760...69N} and \citealt{2014A&A...571A..83P}). More recently, multizone models have been developed to try to understand the stochastic nature of the variability and polarisation (e.g. \citealt{2014ApJ...780...87M} and \citealt{2016ApJ...829...69Z}), however, the spatial flow of electrons and total energy conservation are often neglected. In this paper we build on our successful multizone inhomogeneous fluid jet emission model. This model has been used to fit with unprecedented accuracy to the entire multiwavelength spectrum of a large sample of 42 quiescent blazar spectra in \citet{2015MNRAS.453.4070P}, including radio observations. In this paper we develop a time-dependent inhomogeneous fluid emission model capable of modelling flaring behaviour.

This model has several distinct advantages over previous work: it allows the accurate calculation of the evolution of the spectrum and wavelength-dependent time lags as the plasma propagates down a realistic large scale jet structure based on VLBI observations of M87 \citep{2012ApJ...745L..28A}; it conserves relativistic energy-momentum and particle number flux along the jet so the different emission regions are related by the physical evolution of the accelerating fluid and are not arbitrary physically disconnected emission regions with fixed radii; it can simulate a flaring event which occurs in addition to the average quiescent jet emission, leading to more accurate and complex behaviour than possible in one-zone models; the extended inhomogeneous nature of the jet model allows the radio emission to be accurately calculated. By developing a time-dependent fluid jet emission model we will be able to determine, for the first time, the time-lag between the high energy emission emitted at small jet radii and the low frequency radio emission produced at much larger jet radii where the plasma becomes optically thin.

In this paper we first introduce and briefly describe our quiescent fluid jet model. In section \ref{section3}, the time-dependent flaring model is explained.  We then calculate the characteristic lightcurves and time evolution of the emitted flaring spectrum for flares occurring at different distances along typical BL Lac and flat spectrum radio quasar (FSRQ) type jets in section \ref{section4}. The properties of radio flares and radio time lags are calculated and explained in section \ref{radiosection}. Finally, we fit the model to the observed spectra and lightcurves of the flare in PKS1502+106 from \citet{2010ApJ...710..810A} in section \ref{section6}.

\begin{figure}
          \centering
          \includegraphics[width=8.5cm]{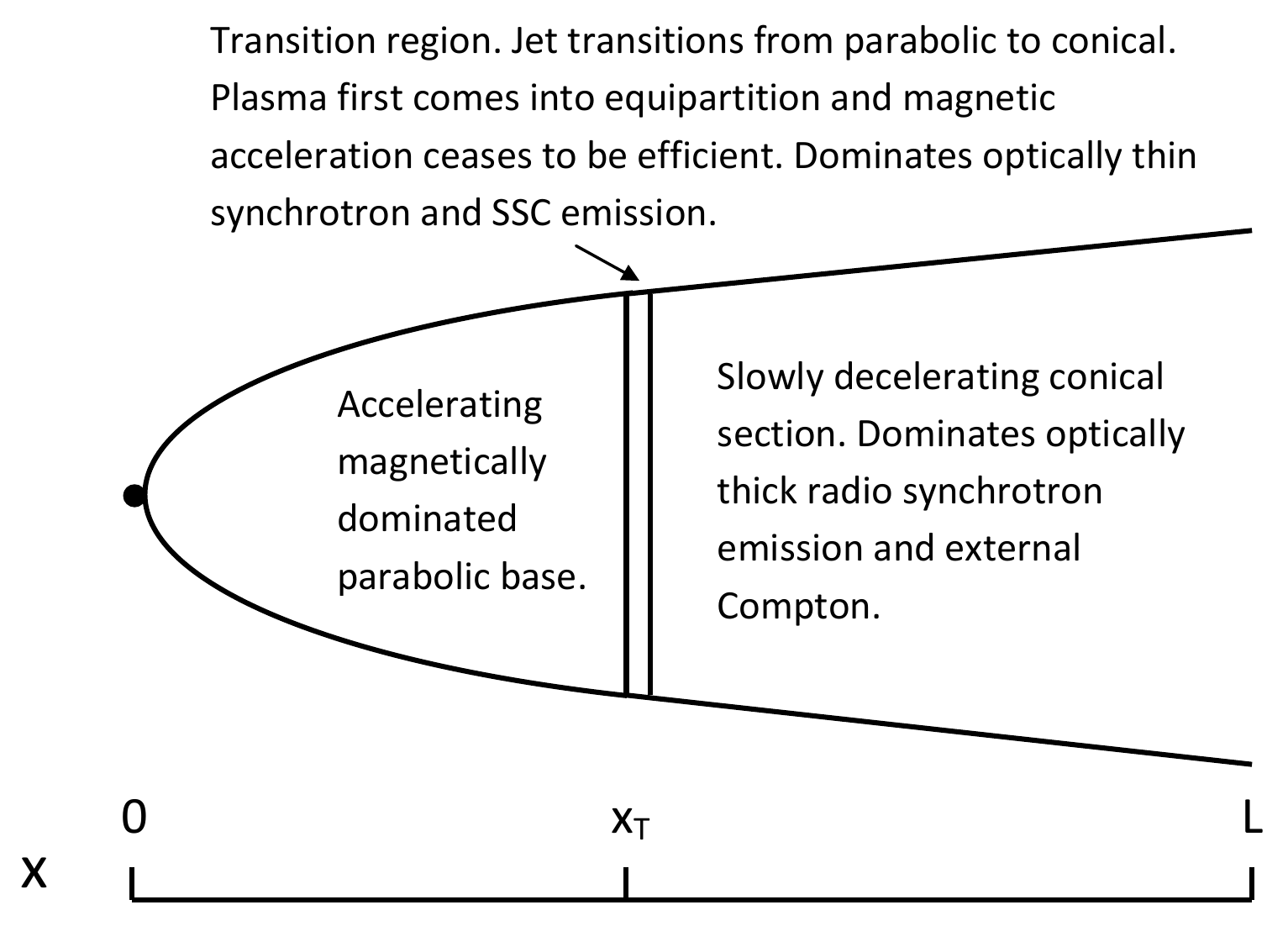}
          \caption{Quiescent jet schematic \citep{2015MNRAS.453.4070P}.}
          \label{schematic}
\end{figure}
\begin{figure}
	\centering
		\includegraphics[width=8cm, clip=true, trim=1.5cm 1cm 2.0cm 1.5cm]{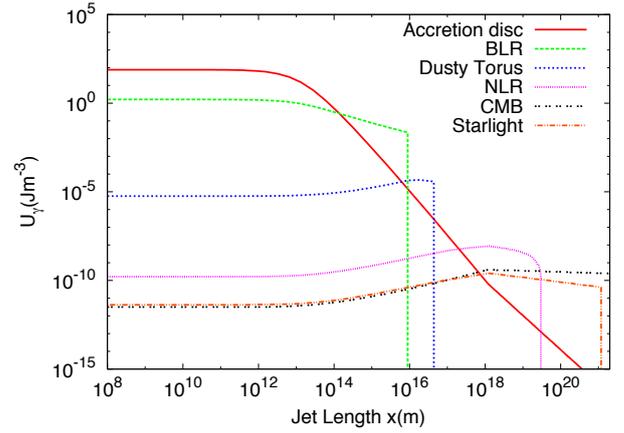} 
                 \caption{The energy density of the different external photon sources as a function of jet length, calculated for the fit to the FSRQ J1512 in the plasma rest frame from \citealt{2015MNRAS.453.4070P}. The rest frame energy densities account for the Doppler boosting and beaming of the external photon sources into the plasma rest frame. The acceleration of the bulk Lorentz factor (up to $x\sim10^{18}$m) of the jet leads to the increase in the energy density of the CMB as measured in the plasma rest frame and the slow deceleration of the conical jet leads to the decrease in the CMB energy density beyond this distance. }
                 \label{Uphot}
\end{figure}

\section{Quiescent jet model} 

Our quiescent jet model was the first inhomogeneous relativistic fluid emission model able to fit the entire blazar spectrum from radio to $\gamma$-rays (the model was developed in a series of papers \citealt{2012MNRAS.423..756P},  \citealt{2013MNRAS.429.1189P}, \citealt{2013MNRAS.431.1840P} and \citealt{2013MNRAS.436..304P}). The model incorporates the latest results from observations and simulations. The jet starts with a parabolic magnetically dominated accelerating base and this transitions to a conical jet which is slowly decelerating due to entrainment once the plasma reaches equipartition (see e.g. \citealt{2006MNRAS.368.1561M} and \citealt{2007MNRAS.380...51K}). This means that the transition of the jet shape between parabolic and conical coincides with the first strong burst of non-thermal particle acceleration in the jet. Non-thermal particle acceleration could be due to either magnetic reconnection or a recollimation shock at the transition region.  The geometrical shape of the jet is given by radio VLBI observations of the jet shape of M87 which is parabolic up to $10^{5}r_{s}$ and thereafter conical \citep{2012ApJ...745L..28A}. As information on the detailed structure of other AGN jets is not as well known, we have chosen our jet geometry to be the same as M87 scaled linearly by an effective black hole mass $M$. Since we do not necessarily expect all jets to have the same shape scaled with black hole mass, the reader should view this effective black hole mass as a fitting parameter. 

The basic equations and assumptions of the model are outlined in the following section, however, for more details the reader should refer to \citealt{2012MNRAS.423..756P} and \citealt{2013MNRAS.429.1189P}. The relativistic fluid model conserves energy-momentum along the jet by solving the standard equation for conservation of relativistic energy-momentum along the jet
\be
\nabla_{\mu}T^{\mu \nu}=0, \qquad T^{\mu \nu}=T^{\mu \nu}_{\m{Magnetic}}+T^{\mu \nu}_{\m{Particles}}+T^{\mu \nu}_{\m{Losses}},
\label{consE}
\ee
where $T^{\mu \nu}$ is the energy-momentum tensor of the jet plasma which has been decomposed into magnetic and particle energy densities, and also a cumulative radiative energy loss term which is required to conserve the total energy along the jet when the plasma suffers radiative and adiabatic energy losses. Under the assumption that the jet plasma is locally isotropic and homogeneous in the fluid rest frame, the rest frame energy-momentum tensor takes the form of a relativistic perfect fluid with energy density $\rho'$ and isotropic pressure $p'=\rho'/3$
\be
T'^{\mu\nu}=\m{diag}(\rho',p',p',p').
\ee
The lab frame energy-momentum tensor can be found by Lorentz transforming the rest frame tensor
\bea
&&T^{\mu \nu}(x)=\Lambda^{\mu}_{\,\,\rho} T'^{\rho \sigma } \Lambda _{\,\,\sigma}^{ \nu}=... \nonumber \\  &&\hspace{-0.7cm} \begin{pmatrix} \gamma_{\m{bulk}}^{2} \rho'(1+\beta_{\m{bulk}}^{2}/3) & \frac{4}{3}\gamma_{\m{bulk}}^{2}\beta_{\m{bulk}} \rho' &0 &0\\\frac{4}{3}\gamma_{\m{bulk}}^{2}\beta_{\m{bulk}} \rho' &\gamma_{\m{bulk}}^{2}\rho'(1/3+\beta_{\m{bulk}}^{2}) &0 &0\\0&0&\frac{\rho'}{3}&0\\0&0&0&\frac{\rho'}{3}\end{pmatrix} \label{LTEM},
\eea 
where $\Lambda$ is the Lorentz transformation tensor, $\gamma_{\m{bulk}}$ is the bulk Lorentz factor of the jet as a function of distance and $\beta_{\m{bulk}}=v_{\m{bulk}}/c$ is the jet velocity divided by the speed of light. We use coordinates in which $x$ is the direction along the jet axis, $R$ is cylindrical radius, $\phi$ azimuthal angle and $t$ is time. Throughout this paper unprimed coordinates are those measured in the lab frame which is at rest with respect to the black hole and primed coordinates to the fluid rest frame, obtained by a Lorentz transformation. Integrating the equation for energy-momentum conservation and using the 4-dimensional divergence theorem (see e.g. \citealt{1974ApJ...191..499P}) we find our governing equation of energy conservation
\bea
&&\hspace{-1.0cm}\int \nabla_{\mu}T^{\mu \nu} d^{4}V=\oint T^{\mu \nu} d^{3}S_{\mu}= \nonumber\\
&&\hspace{-0.7cm}\left[\int_{x}^{x+\Delta x}\int_{0}^{2\pi}\int_{0}^{R} T^{t\nu} RdRd\phi dx\right]_{t}^{t+\Delta t} \nonumber\\
&&\hspace{-1.0cm}+\left[\int_{t}^{t+\Delta t}\int_{0}^{2\pi}\int_{0}^{R} T^{x\nu} RdRd\phi dt\right]_{x}^{x+\Delta x}\nonumber\\ &&\hspace{-1.0cm}+\left[\int_{t}^{t+\Delta t}\int_{x}^{x+\Delta x}\int_{0}^{2\pi} T^{R\nu} Rd\phi dxdt\right]_{0}^{R} \nonumber\\ &&\hspace{-1.0cm}+\left[\int_{t}^{t+\Delta t}\int_{x}^{x+\Delta x}\int_{0}^{R} T^{\phi\nu} dRdxdt\right]_{0}^{2\pi}=0,\nonumber\\ \label{divE}
\eea
where the invariant 4-volume, $d^{4}V=\sqrt{|g|}dxdRd\phi dt$, and $g$ is the determinant of the metric. Since the majority of jet emission occurs at a considerable distance away from the black hole, where the effects of general relativity become unimportant, we can approximate $\sqrt{|g|}=R$. Due to the assumed time-independence and homogeneity perpendicular to the jet axis the first and fourth brackets vanish. Since the jet fluid is confined within its radius R(x), their is no energy-momentum flux through the outer radial jet boundary (except radiative losses which are included separately via $T^{\mu\nu}_{\m{losses}}$) so the third bracket also vanishes. Expanding the second bracket we find
\be
\left[T^{x\nu}\pi R^{2}\Delta t\right]_{x}^{x+\Delta x}=\frac{\partial}{\partial x}\left(T^{x\nu}\pi R^{2}\Delta t\right)\Delta x=0.
\ee
The only non-trivial equations occur when $\nu=t, x$ and they lead to the same equation for conservation of energy-momentum in the relativistic limit, $\beta_{\m{bulk}}\rightarrow 1$
\be
\frac{\partial}{\partial x}\left(\frac{4}{3}\gamma_{\m{bulk}}(x)^{2}\pi R^{2}(x)\rho'(x)\right)=0.
\label{ce}
\ee
This equation ensures conservation of total energy (including radiative energy losses) along the jet. Particle number flux is conserved along the jet by the usual equation
\be
\nabla_{\mu}J_{e}^{\mu}(E_{e},x)=0, \qquad J_{e}^{\mu}(E_{e})=n'_{e}(E_{e},x)U^{\mu}(x).
\label{consJ}
\ee
where $U^{\mu}(x)=\gamma(x)(1,\beta(x),0,0)$ is the jet fluid 4-velocity, $\beta(x)=v(x)/c$, $v(x)$ is the jet speed and $n'_{e}$ the electron (and positron) number density in the rest frame. Integrating (\ref{consJ}) and using the divergence theorem as before we find
\be
\frac{\partial}{\partial x}[\pi R^{2}(x)n'_{\m{e}}(x)U^{0}(x)]=0,
\label{cc}
\ee
The emitted radiation of a population of non-thermal electron-positron pairs due to synchrotron and inverse-Compton scattering is calculated by dividing the extended jet structure into thousands of cylindrical sections with adaptive widths. The widths are chosen to resolve both the shortest radiative lengthscale and any change in the jet radius of more than $10\%$. Hereafter we shall use electron to refer to both electrons and positrons. The inverse-Compton emission is calculated by integrating the full Klein-Nishina cross-section in the plasma rest frame. All relevant inverse-Compton seed photon sources are calculated accurately as a function of distance along the jet including: synchrotron self-Compton (SSC), direct accretion disc, broad line region (BLR), dusty torus, narrow line region (NLR), starlight and the cosmic microwave background (CMB), as shown in Fig \ref{Uphot} (see \citealt{2013MNRAS.429.1189P} for details). The synchrotron photons emitted by neighbouring sections are also included when calculating the SSC emission from each individual section (cross-zone SSC), although we find this effect to be negligible. The line of sight optical depth due to synchrotron self-absorption and photon-photon pair production is also calculated for each section along the jet (for more details see \cite{2012MNRAS.423..756P} and \cite{2013MNRAS.436..304P} respectively). 

The model is leptonic and assumes that the relativistic electron-positron jet plasma is generated by pair production at the jet base by a spark gap mechanism operating in the magnetosphere (\citealt{1969ApJ...157..869G}, \citealt{1977MNRAS.179..433B} and \citealt{2015ApJ...809...97B}). Further along the jet when particle acceleration occurs, we assume a freshly accelerated non-thermal spectrum given by a power-law and exponential cutoff (typical for both magnetic reconnection and shock acceleration processes e.g. \citealt{2001ApJ...562L..63Z}, \citealt{2014ApJ...783L..21S} and \citealt{2011MNRAS.tmp.1506B} and \citealt{2012ApJ...745...63S}).
\be
N_{\m{injected}}(x,E_{e})=AE_{e}^{-\alpha} e^{-E_{e}/E_{\m{max}}}, \label{Ch2loss2}
\ee
where $\alpha$ is the electron energy distribution spectral index. These freshly accelerated electrons are then added to the existing non-thermal electron population. The radiative energy losses due to synchrotron and inverse-Compton radiation emitted by the non-thermal electrons are calculated self-consistently, changing the electron energy spectrum as the plasma travels along the jet.
\begin{figure}
          \centering
          \includegraphics[width=8.5cm, clip=true, trim= 4cm 16cm 5cm 1cm]{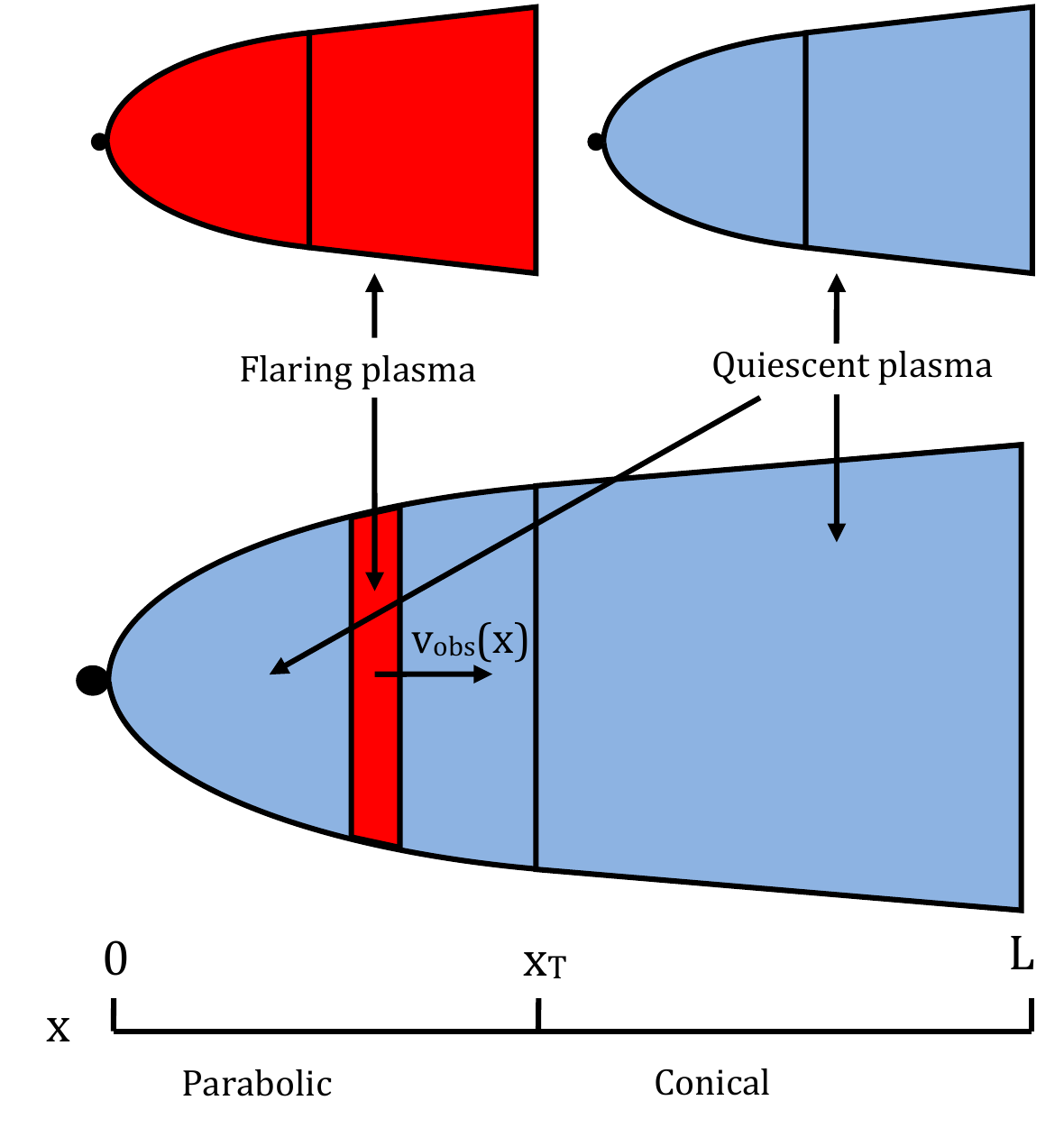}
          \caption{Schematic of our flaring model. The evolution and emission of the quiescent and flaring plasma jets are first calculated separately by propagating emitting plasma with different properties along the jet conserving energy-momentum and particle number. The time-dependent flare emission is calculated by splicing the relevant length of flaring plasma into the quiescent jet and integrating the total time-dependent line-of-sight synchrotron and pair-production opacities through the jet. The evolution of the location and length of flaring plasma depends upon the Doppler-boosted velocity and duration of the flare as measured by an observer (\ref{xflare}). The corresponding location and width of the flaring section which is spliced into the quiescent model is therefore time-dependent.}
          \label{schematic2}
\end{figure}

\section{Flaring jet model} \label{section3}

It is relatively straightforward to create a time-dependent flaring jet model building upon our existing quiescent fluid jet model. The quiescent jet model calculates the emission and radiative energy losses for a thin slab of plasma propagating along the entire jet. Using the assumed time-independence of the quiescent jet, the emission from the jet at a particular distance will be unchanging and will therefore have the same plasma properties that the propagating thin slab possessed when it was at that location. In order to calculate the time-dependent emission from a section of plasma with different initial properties to the quiescent plasma, the code is run once to calculate the propagation and emission of the quiescent plasma along the jet and once to calculate the propagation and emission of the \lq{}flaring\rq{} plasma along the jet. The time-dependent solution can then be calculated by choosing an initial width of \lq{}flaring\rq{} plasma at the jet base and then propagating this along the quiescent jet with the correct time-dependent Doppler boosted velocity as measured by an observer (\ref{xflare}). The composite spectrum is calculated by summing the total emission from the quiescent and \lq{}flaring\rq{} sections of the jet and including the integrated line-of-sight synchrotron and pair production opacities through the composite jet. This process is illustrated in Figure \ref{schematic2}.

This method is very efficient at computing time-dependent spectra since it only takes twice as long as the single calculation of a quiescent spectrum, which typically takes $\sim$1 minute. This is because the time taken to splice the flaring plasma into the quiescent jet and sum the total emission is negligible in comparison to the time taken to calculate the emission and energy losses of the propagating plasma along the jet. It is important to note that despite the apparent simplicity and efficiency of this numerical method, the physical accuracy of the model is not compromised. One of the primary purposes of this paper is to investigate the spectral signatures of particle acceleration events which occur at different distances along the jet, particularly in the parabolic base. This is easily achieved using this method by specifying a strong burst of non-thermal particle acceleration to occur within the parabolic base in the flaring jet. More complex flaring events can also be modelled but at extra computational cost. For example, a flaring event at a fixed location in which the energy injected into accelerating non-thermal electrons is time-dependent, smoothly increasing from zero to a maximum and decreasing back to zero. This can be discretised and modelled by solving the propagation and emission for multiple flaring plasma jets, each experiencing different incremental amounts of particle acceleration at the location of the flare. In general, this will clearly be required to model observed flares (and is demonstrated in section \ref{section6}), in which the amplitude of energy injected into particle acceleration by the flaring event will not be well described by a rectangular or \lq{}top-hat\rq{} function in time. 

To calculate the time-dependent spectrum produced by a flaring section of plasma propagating along the quiescent jet it is useful to work in observer quantities. We define the initial width of the flaring section in terms of the observed length of time of the flare, $\Delta t_{\m{obs}}$. This is the observed duration that the flaring section of plasma is injected at the base of the jet. The observer will measure velocities of the plasma which have been Doppler boosted and so the observed lab frame distance of the plasma will be given by
\be
x(t_{\m{obs}}+dt_{\m{obs}})=x(t_{\m{obs}})+v(x)dt_{\m{obs}}\frac{\delta_{\m{Doppler}}(x)}{1+z},
\label{xflare}
\ee
\be
\delta_{\m{Doppler}}=\frac{1}{\gamma_{\m{bulk}}(1-\beta_{\m{bulk}}\cos\theta_{\m{obs}})},
\ee
where $z$ the cosmological redshift, $\delta_{\m{Doppler}}$ is the observed Doppler factor and $\theta_{\m{obs}}$ the angle between the jet axis and the observer. The advantage of working with observed quantities is that the observed separation in time, between the start and end of the flaring section passing through the same point in the lab frame, is constant for the observer. We define this time as $\Delta t_{\m{flare}}$. Since the observed velocities at the start and end points of the flaring section will not in general be equal, the observed width of the flaring section will change as a function of time. A flaring section with observed duration $\Delta t_{\m{flare}}$, corresponds to a section of flaring plasma with start and end points at $x_{1}$ and $x_{2}$ respectively. The evolution of the starting location $x_{1}(t_{\m{obs}})$ is given by (\ref{xflare}) and $x_{2}(t_{\m{obs}})$ is simply $x_{2}(t_{\m{obs}})=x_{1}(t_{\m{obs}}-\Delta t_{\m{flare}})$. This leads to the standard relativistic result that flares appear shorter and more luminous to an observer looking down the jet compared to the flare in the lab frame.

The distance of the transient flaring front, at which the flaring plasma first experiences particle acceleration, is defined as $x_{\m{flare}}$. The duration of the particle acceleration occurring at the front is simply the observed duration of the flare, $\Delta t_{\m{flare}}$. The time-dependent power injected into non-thermal particle acceleration is quantified by, $\epsilon_{\m{flare}}(t)$, the equipartition fraction of the plasma on immediately passing through the flaring front. The equipartition fraction, or magnetisation, is the ratio of energy density in non-thermal particles to the magnetic energy density i.e. $\epsilon_{\m{flare}}=U_{e\pm}/U_{B}$. It is worth noting that since total energy is conserved in our fluid jet model, the energy which goes into the acceleration of non-thermal particles must be converted from either the existing magnetic or bulk kinetic energy of the jet plasma. In the magnetically dominated accelerating parabolic base any energy converted into the acceleration of non-thermal electrons is assumed to originate from magnetic energy via magnetic reconnection.       

\subsection{Light-travel time constraints}

One of the biggest problems encountered when modelling flaring events is to reconcile the short duration of the flare with the large physical volume of plasma required to emit sufficient energy to produce the flare (e.g. \citealt{2007ApJ...669..862A} and \citealt{2007ApJ...664L..71A}. Most particle acceleration mechanisms originate from causally-connected physical processes e.g. the energy extracted from magnetic reconnection proceeds at a rate determined by the velocity of magnetised plasma entering the low pressure reconnection region. This is intrinsically a causally-connected process since the plasma must receive a signal informing it of the developing low pressure region in order to be able to accelerate towards it. Of course, it is also possible that a stochastic flaring mechanism, based on the random coincident emission from multiple disconnected regions could provide a mechanism for powerful flares. The flaring from these emission regions would not have to be causally-connected and so could thereby occur on shorter timescales than the light travel time across the region. In this paper we shall assume the particle acceleration mechanism is a causally-connected process such as a large-scale shock or reconnection event. We quantify whether the flaring region in our model is causally-connected by introducing the parameter $\Delta_{\m{flare}}=\Delta t_{\m{flare}}/\Delta t_{\m{lc}}$, the ratio of the observed flaring timescale to the observed light-crossing timescale of the flaring front
\be
\Delta_{\m{flare}}=\frac{\Delta t_{\m{flare}}}{\Delta t_{\m{lc}}}=\frac{\Delta t_{\m{flare}}\delta_{\m{Doppler}}c}{(1+z)R}. 
\label{Deltaflare}
\ee
where in the observer frame the light-crossing timescale is decreased by cosmological redshift $(1+z)$ and increased by Doppler boosting $\delta_{\m{Doppler}}$. This parameter essentially quantifies how elongated (or \lq{}pancake-like\rq{}) the region of flaring plasma is i.e. the ratio of its width along the jet axis to its cylindrical radius. For a flaring event to be causally-connected we require $\Delta_{\m{flare}}\geq1$, which is satisfied by all the flaring events modelled in this paper.  

\subsection{Geometrical time delay and shape of the flaring front}\label{geometricaldelay}

The flaring front is assumed to be perpendicular to the jet axis in this work. This is chosen because of the 1D nature of the model, and  because the detailed structure of the particle acceleration front is not yet known from observations. The expected theoretical shape of the front depends upon the particle acceleration mechanism and would likely take the form of a Mach-cone in the case of particle acceleration by a recollimation shock, or the reconnection surface if particles are accelerated by case of magnetic reconnection. The shape and time-dependent propagation of the particle acceleration front affects the shape of the lightcurves by introducing time delays in the observed radiation. These time delays have both geometrical and physical components. The geometrical component arises because of the differences in the observed path length across the flaring front (we shall calculate this for our model below). The physical time delay arises from differences in the time at which particle acceleration occurs at different points along the flaring front i.e. the time taken for a recollimation shock front to propagate through the jet interior.  The geometrical and physical delays will cause a smoothing of the lightcurve on a timescale given by the combined geometrical and physical delays, typically of order $\Delta t\sim R/(\Gamma_{\m{Doppler}}c)$, comparable to the causal connection timescale of the jet.

Let us now calculate the geometrical time delay in our model. The geometrical delay is caused by the path length difference between the observer and different points along the flaring front. These differences in path length result in an increased light travel time at different points along the front given by
\be
\Delta t_{g}(r)=\frac{r\sin\theta_{\m{obs}}}{c},\label{geometrictimescale}
\ee
where $r$ is the radial jet coordinate in the direction of the component of the observer's line of sight which is perpendicular to the jet axis. To obtain our observed lightcurve and spectra we need to include this observed time delay for radiation emitted at different points along the flaring front. To calculate the observed radiation flux it is necessary to take into account the emitting surface area of the jet as a function of $r$.  
\be
F_{\nu}(t)=\frac{\int_{-R}^{R} F_{\nu}(t+\Delta t_{g})2\sqrt{R^{2}-r^{2}}dr}{\pi R^{2}},\label{geom}
\ee         
where $2\sqrt{R^{2}-r^{2}}dr=dA$ is the area element of the effective emitting surface, as a function of $r$ and $\pi R^{2}$ is the normalisation (total effective emitting surface). Clearly since the model presented here is 1D, effects which intrinsically depend on the radial jet structure can only be taken into account approximately. However, we do not expect our results to differ substantially from a 2D/3D model since the effect of the physical and geometrical time delays is to smooth the spectrum on a timescale $\sim R\sin\theta_{\m{obs}}/c\sim R/(\gamma c)$, (since blazars are observed close to the line of sight and Doppler boosted so $\theta_{\m{obs}}\sim 1/\gamma_{\m{bulk}}$). These delays act to effectively smooth over radial structure, making a 1D model, such as the one presented here, a good first approximation.    

\begin{figure*}
	\centering
		\subfloat[$x_{\m{flare}}=3.04\times10^{-5}\m{pc}$,\qquad $\Delta t_{\m{flare}}=$4.76 minutes]{ \includegraphics[width=7cm, clip=true, trim=0.0cm 0cm 0.cm 0cm]{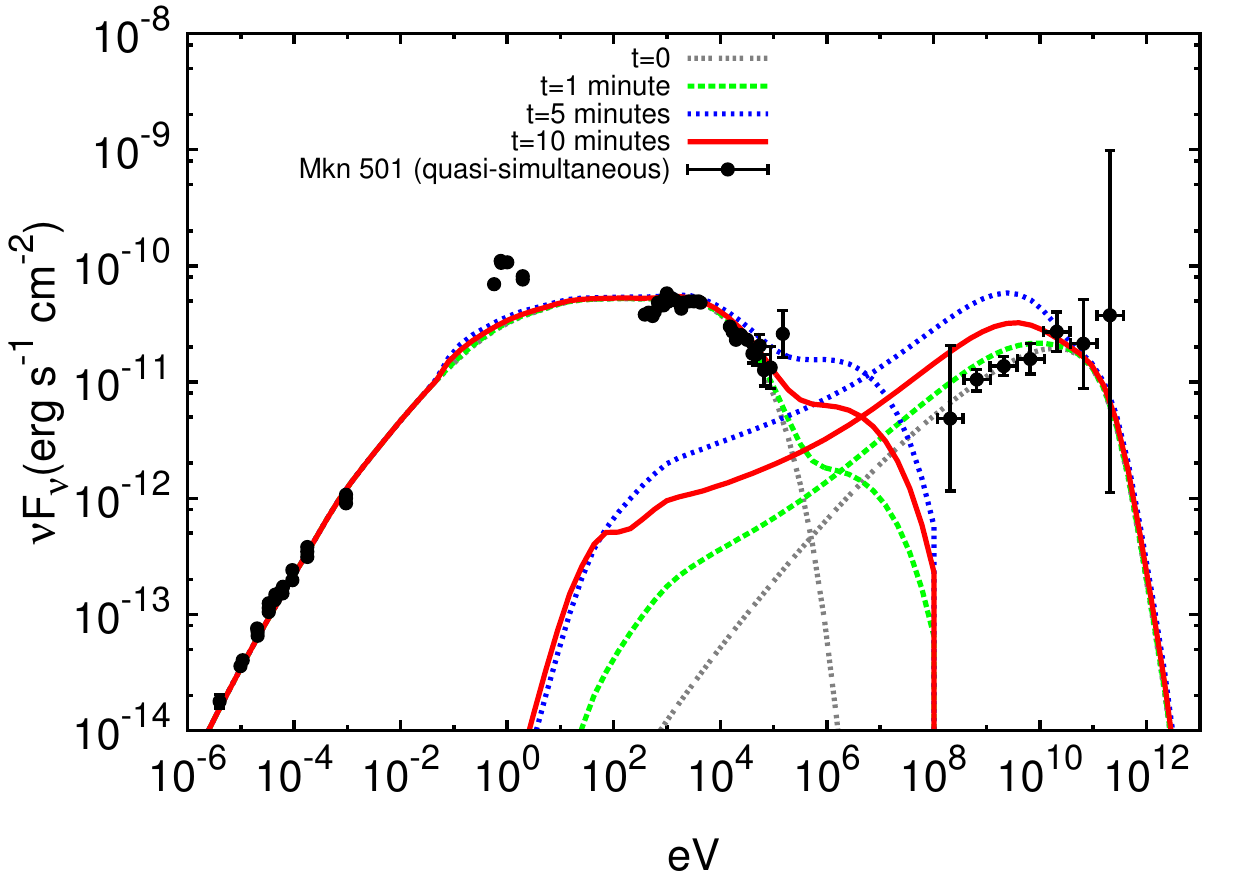} }
\qquad
		\subfloat[$x_{\m{flare}}=3.04\times10^{-5}\m{pc}$,\qquad $\Delta t_{\m{flare}}=$4.76 minutes]{ \includegraphics[width=7cm, clip=true, trim=0cm 0cm 0.cm 0cm]{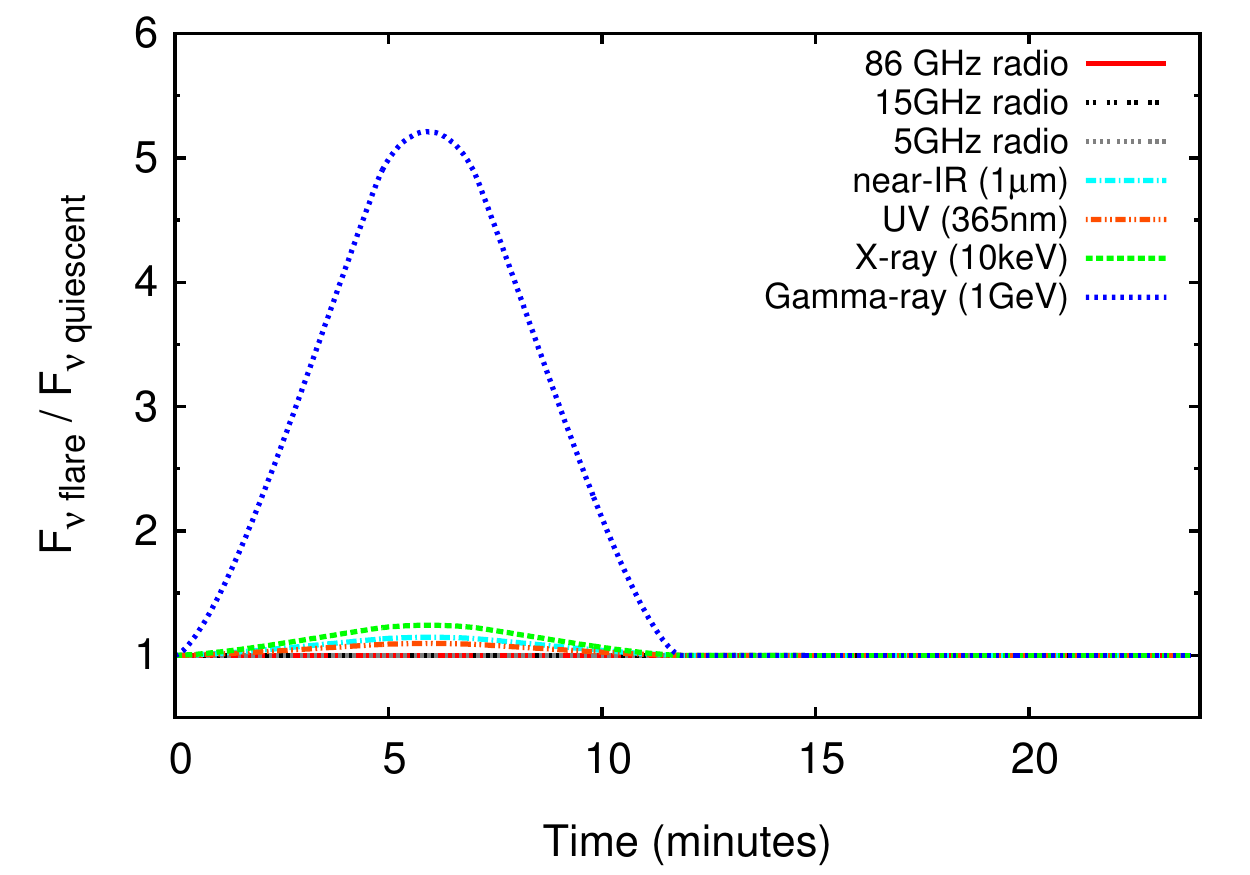} } 
\vspace{-0.2cm}
		\subfloat[$x_{\m{flare}}=3.04\times10^{-4}\m{pc}$,\qquad $\Delta t_{\m{flare}}=$16.4 minutes]{ \includegraphics[width=7cm, clip=true, trim=0cm 0cm 0.cm 0cm]{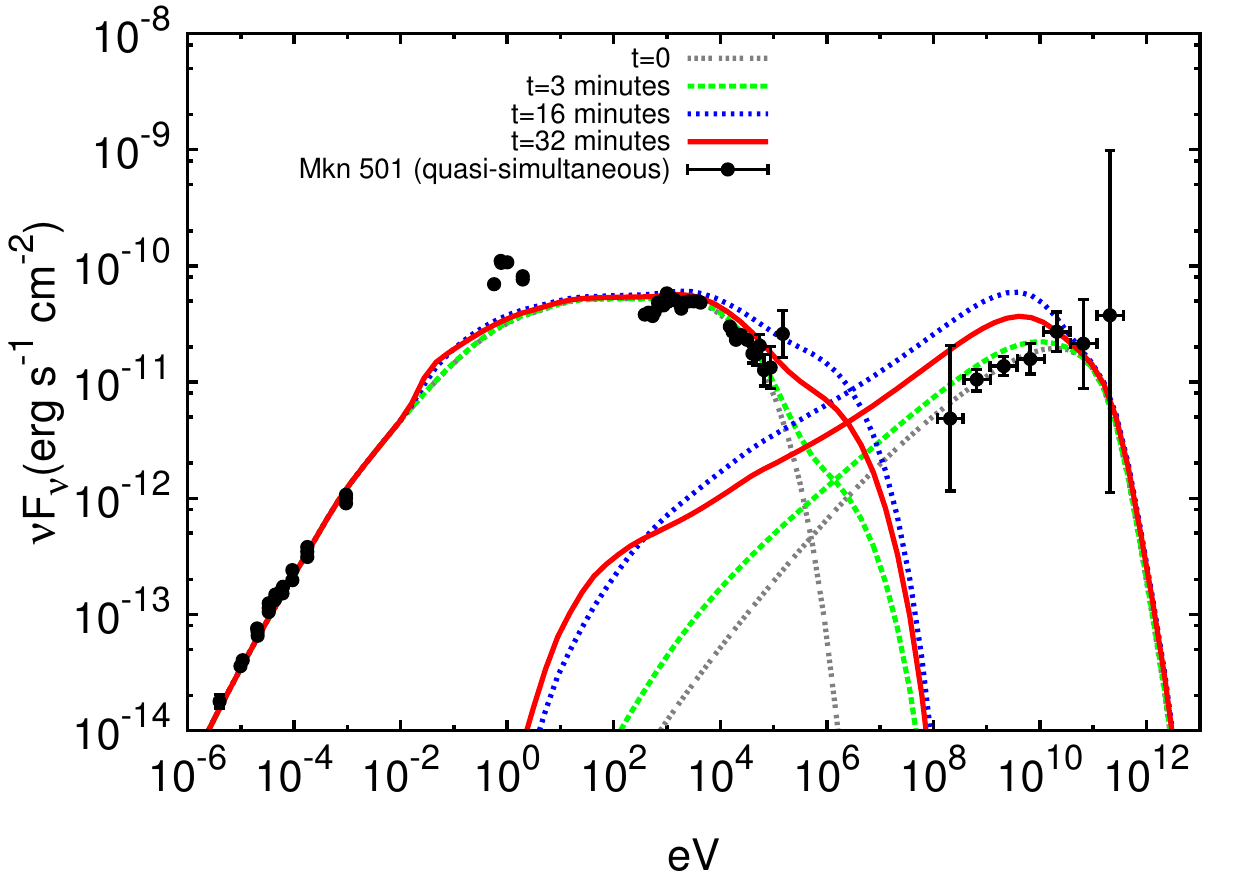} }
\qquad
		\subfloat[$x_{\m{flare}}=3.04\times10^{-4}\m{pc}$,\qquad $\Delta t_{\m{flare}}=$16.4 minutes]{ \includegraphics[width=7cm, clip=true, trim=0cm 0cm 0.cm 0cm]{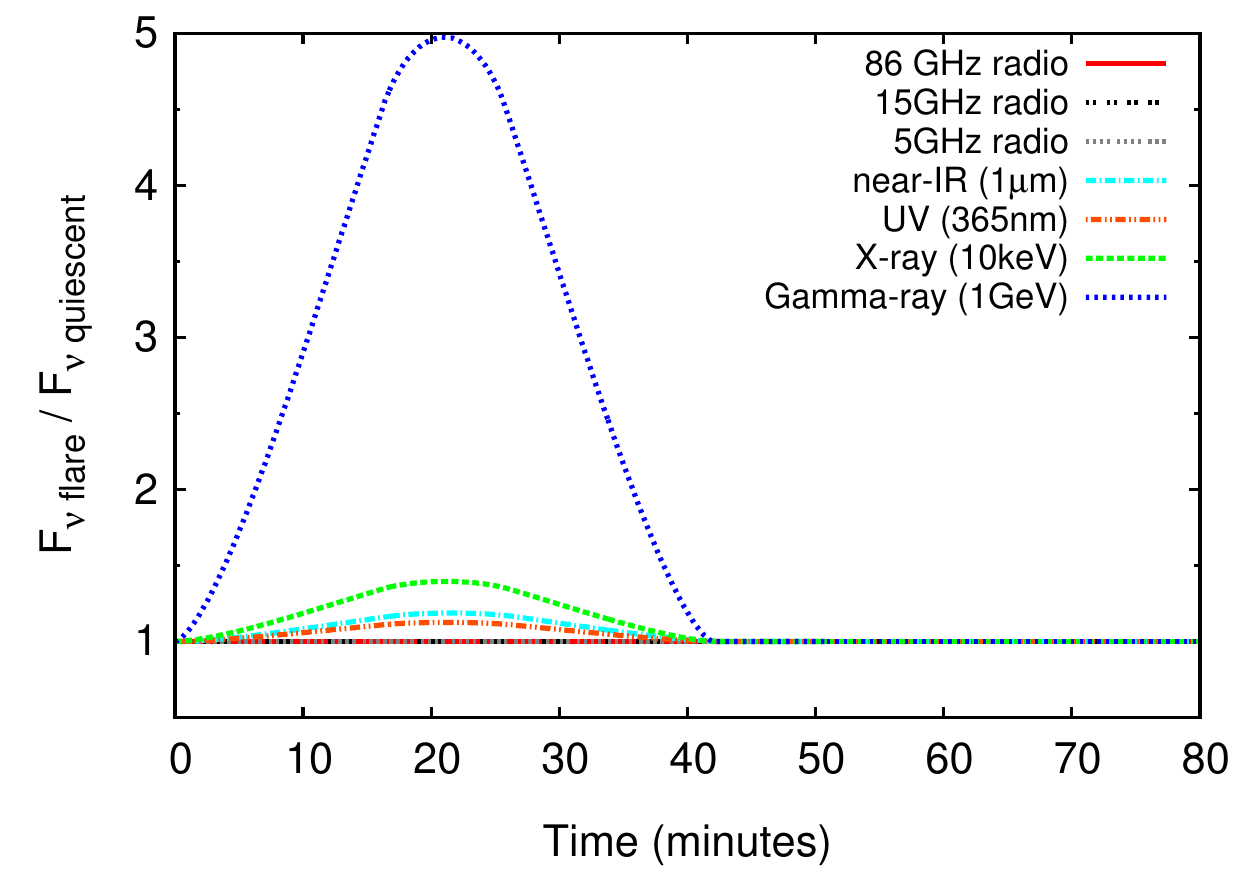} }
\vspace{-0.2cm}
\subfloat[$x_{\m{flare}}=3.04\times10^{-3}\m{pc}$,\qquad $\Delta t_{\m{flare}}=$57.5 minutes]{ \includegraphics[width=7cm, clip=true, trim=0cm 0cm 0.cm 0cm]{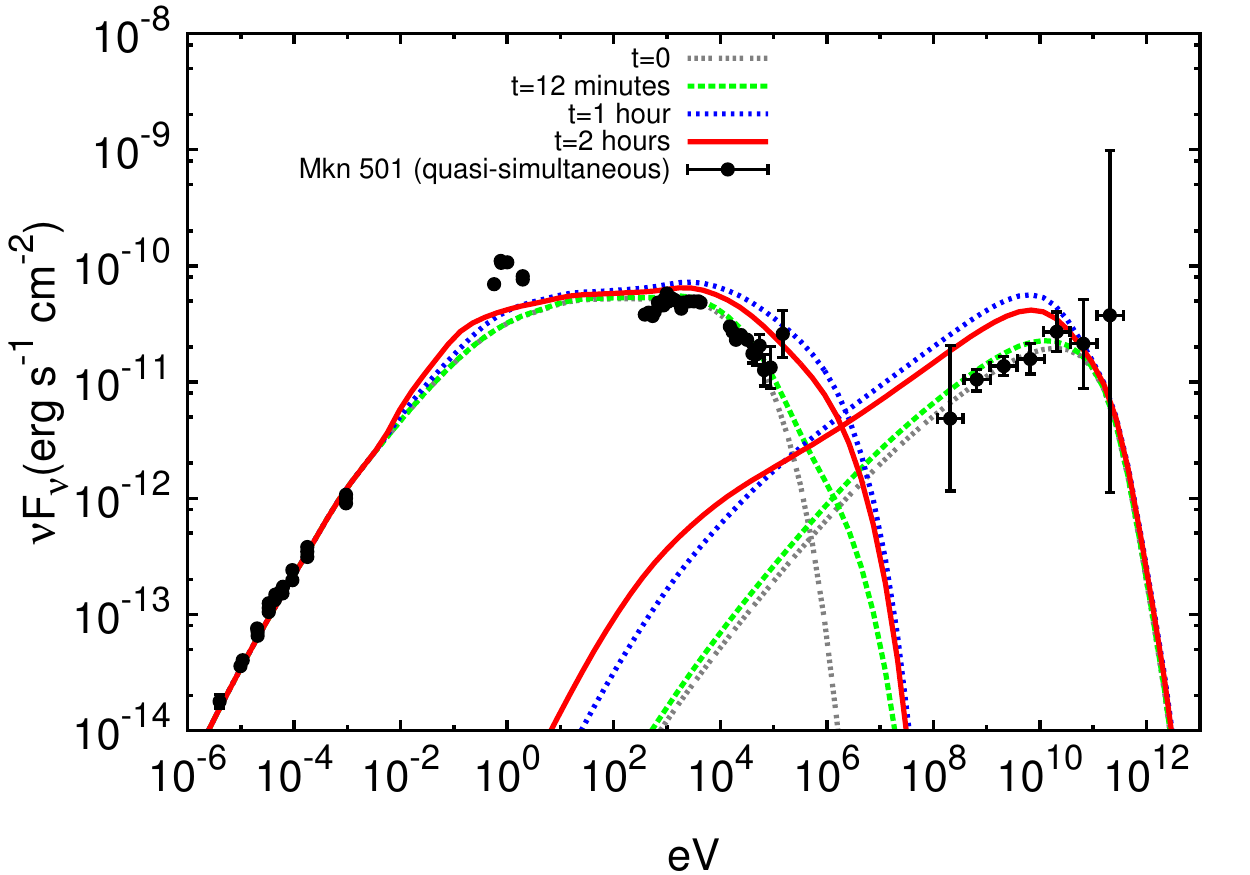} }
\qquad
		\subfloat[$x_{\m{flare}}=3.04\times10^{-3}\m{pc}$,\qquad $\Delta t_{\m{flare}}=$57.5 minutes]{ \includegraphics[width=7cm, clip=true, trim=0cm 0cm 0.cm 0cm]{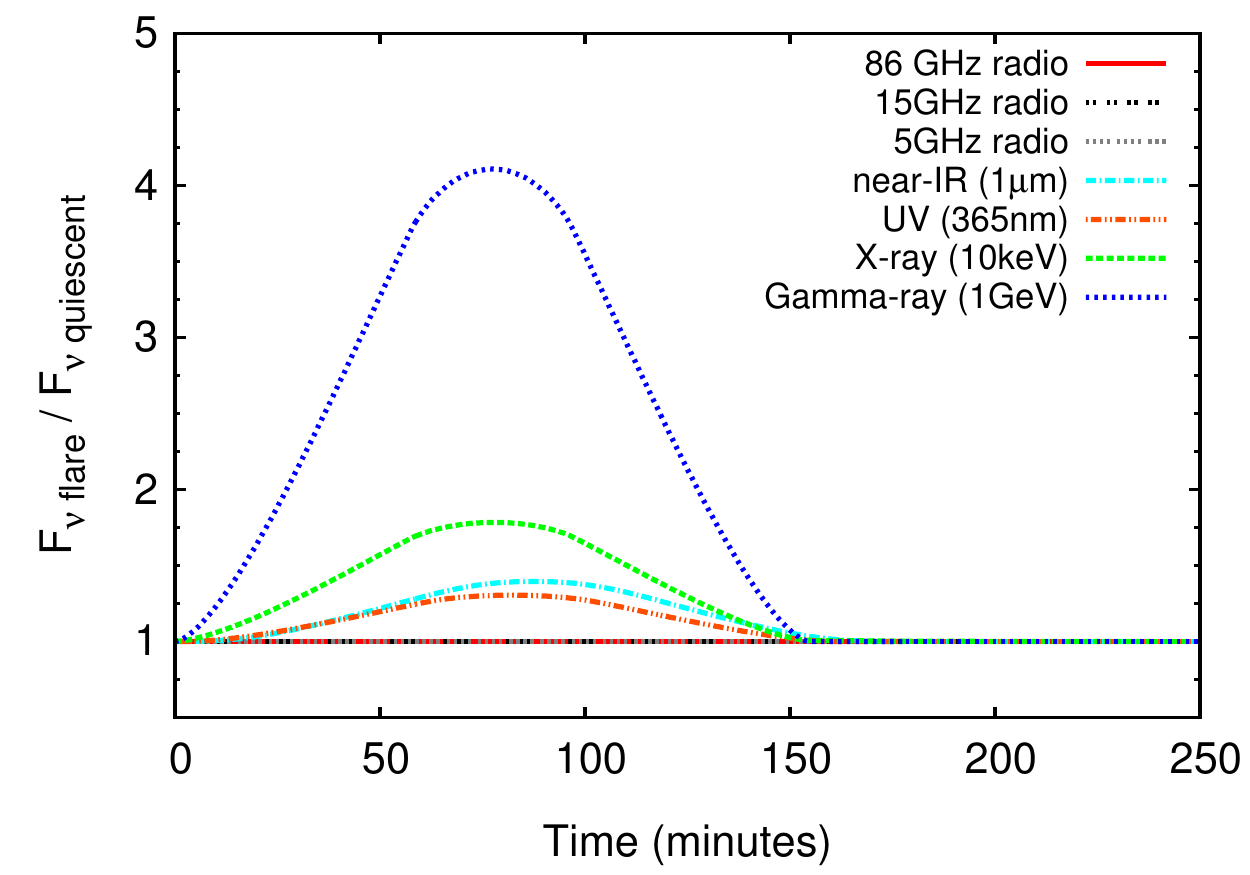} }
\vspace{-0.2cm}
		\subfloat[$x_{\m{flare}}=0.0304\m{pc}$,\qquad $\Delta t_{\m{flare}}=$3.33 hours]{ \includegraphics[width=7cm, clip=true, trim=0cm 0cm 0.cm 0cm]{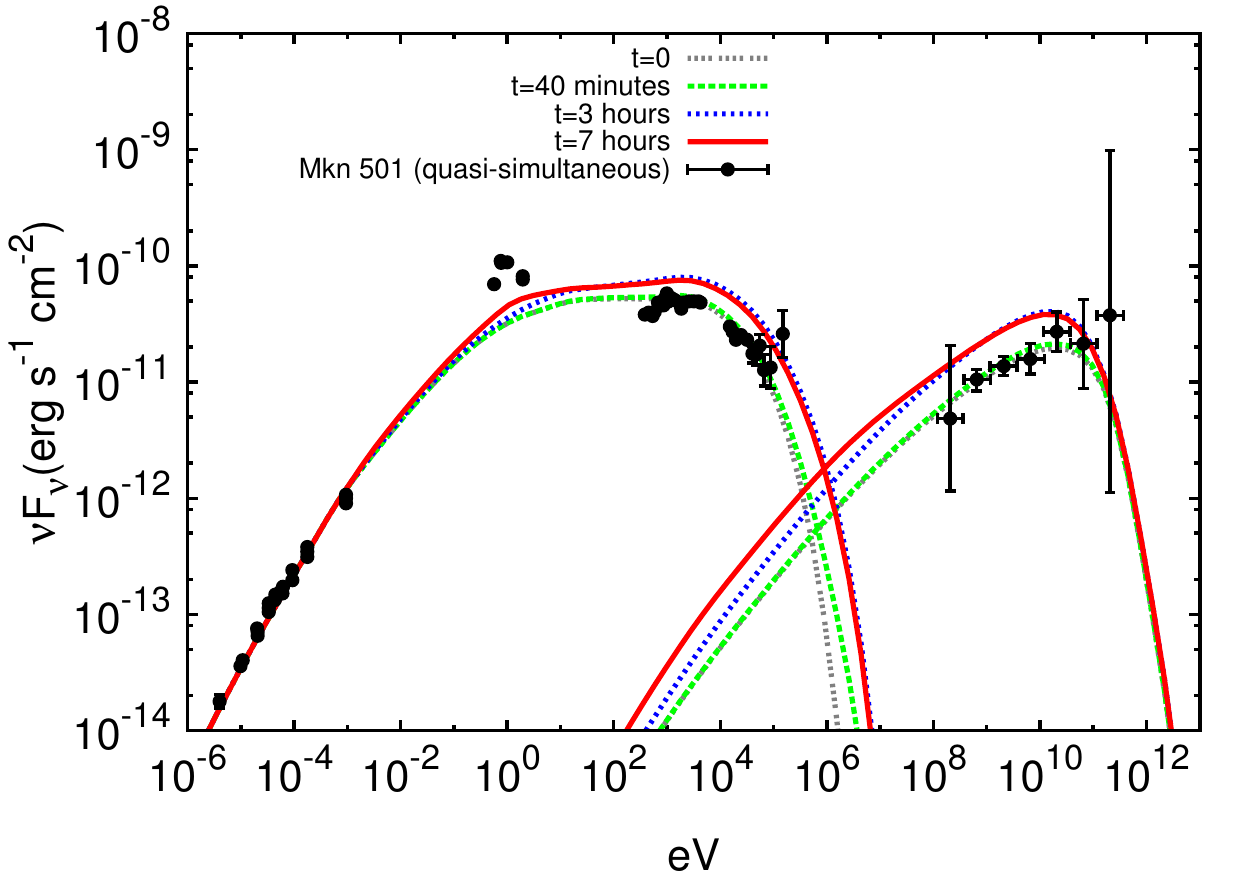} }
\qquad
		\subfloat[$x_{\m{flare}}=0.0304\m{pc}$,\qquad $\Delta t_{\m{flare}}=$3.33 hours]{ \includegraphics[width=7cm, clip=true, trim=0cm 0cm 0.cm 0cm]{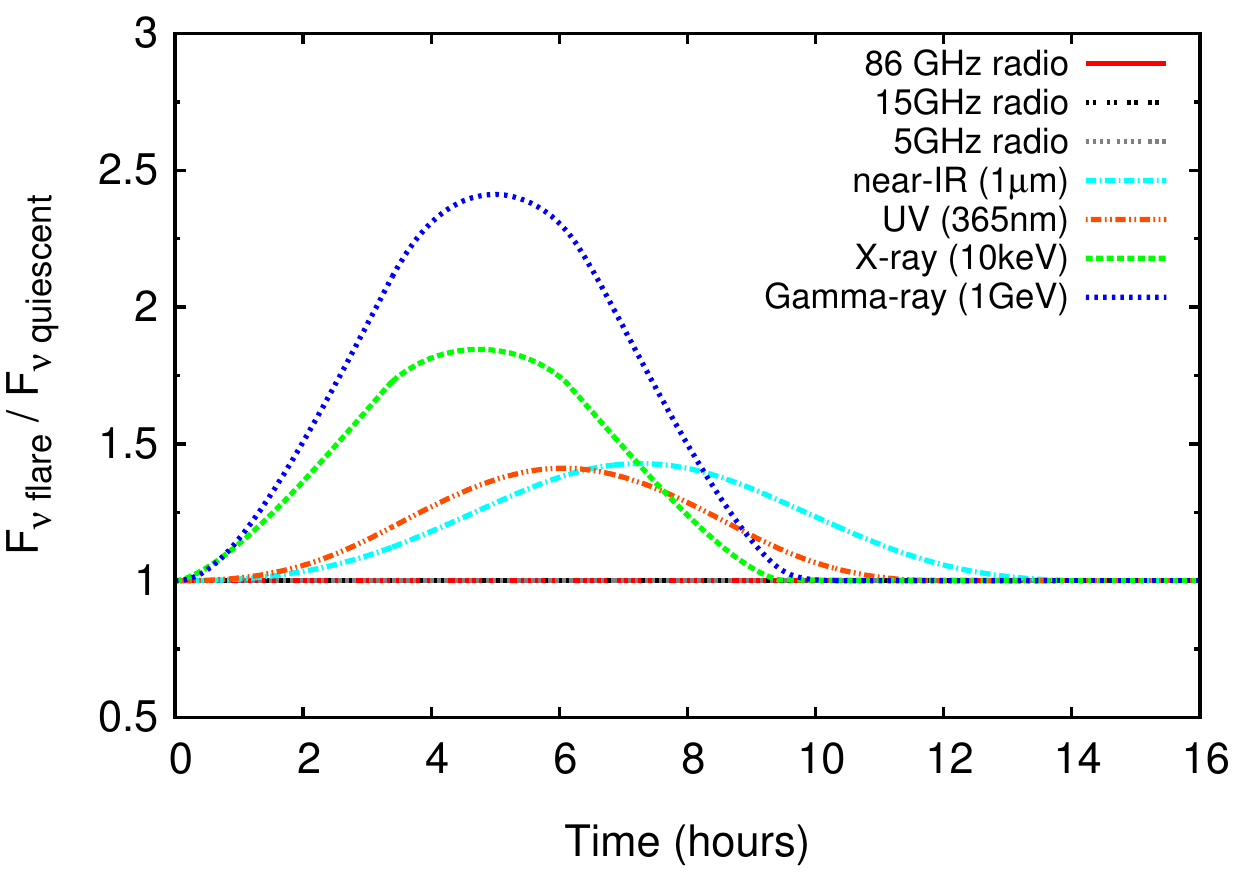} }	
	\caption{The evolution of the spectrum and lightcurves caused by flaring events at different locations along the parabolic base of the jet in Markarian 501 which occur in addition to the quiescent jet emission (the transition between parabolic and conical occurs at $0.304\m{pc}$). Flares closer to the jet base are more luminous with higher Compton-dominance due to the larger magnetic field strengths and smaller jet radius. The flaring rise and decay timescales are approximately equal to the radiative lifetime of the electrons emitting at that frequency smoothed by the time-delay caused by differences in the geometrical path length across the flaring front (see text and Fig. \ref{schematic3} for discussion).} 
\label{BLflare1}
\end{figure*}

\begin{table*}
\centering
\begin{tabular}{| c | c | c | c | c | c | c | c |}
\hline
Blazar & Figure no. & $x_{\m{flare}}$ & $\Delta t_{\m{flare}}$ & $\Delta_{\m{flare}}$ & $\epsilon_{\m{flare}}$ & $W_{J\m{q}}/W_{J\m{f}}$ & $E_{\m{max\ f}}/E_{\m{max\ q}}$ \\ \hline
Markarian 501 & 4 & variable & variable & 1 & 1 & 1 & 1 \\ \hline
PKS1510-089 & 6 & variable & variable & 1 & 1 & 1 & 1 \\ \hline
\vspace{-0.3cm} & & & & & & \\
Markarian 501 & 8 & 0.304pc ($10^{5}r_{\m{s}}$) & 10days, $10^{2}$days & 20, 200 & 1 & 10 & 1 \\ \hline
\vspace{-0.3cm} & & & & & & \\
PKS1510-089 & 8 & 40.2pc ($10^{5}r_{\m{s}}$) & $10^{3}$days, $10^{4}$days & 72, 720 & 1 & 10 & 1 \\ \hline
PKS1502+106  & 9 & 0.312pc (1070$r_{\m{s}}$) & 7 days & 2.7 & variable & 1.45 & 0.537 \\ \hline

\end{tabular}
\caption{Table of flaring parameters for the calculations in this paper. The parameters are the location of the flaring front $x_{\m{flare}}$, the observed duration of the flare $\Delta t_{\m{flare}}$, the causal-connection parameter of the flare $\Delta_{\m{flare}}$ (\ref{Deltaflare}), the equipartition fraction of non-thermal particle to magnetic energy of plasma on immediately passing through the flaring front $\epsilon_{\m{flare}}$, the ratio of the jet power of the flaring plasma ($W_{J\m{f}}$) to the quiescent jet power ($W_{J\m{q}}$), and the ratio of the maximum electron energy injected into the flaring and quiescent plasma $E_{\m{max\ f}}/E_{\m{max\ q}}$ (see equation \ref{Ch2loss2}). Quiescent jet parameters are shown in Table \ref{Table2}.   }
\label{Table1}
\end{table*}

\section{Characterising flares at different distances along the jet} \label{section4}

Our purpose now is to use this model to understand how the location of a flaring event along the jet affects the observed time-dependent spectrum and lightcurve of the flare. This is crucial if we want to determine the location of the flaring region in blazars, one of the most important current questions in the field. We consider flaring events which occur at a range of distances along the parabolic jet base. The flaring event is modelled as a temporary flaring front at a fixed spatial location which accelerates non-thermal electrons in the plasma passing through the front. To simplify the situation, in this section the amount of particle acceleration at the flaring front is chosen so that the plasma will be in equipartition on immediately passing through the front i.e. $\epsilon_{\m{flare}}=1$, for the duration of the flare. The initial properties of the plasma passing through the flaring front will be identical to the quiescent plasma making up the rest of the jet. The properties of this quiescent plasma have been determined by fitting the quiescent multiwavelength spectra of the blazars \citep{2015MNRAS.453.4070P}. In this regard the only difference between the properties of the flaring and quiescent jet plasma is that the flaring plasma experiences particle acceleration at the flaring front, in addition to the particle acceleration which occurs in the quiescent jet. The duration of particle acceleration at the flaring front is chosen such that the flaring region is in causal contact with itself i.e. $\Delta_{\m{flare}}=1$. Let us now consider the effect of flaring events occurring at a variety of distances along BL Lac and FSRQ type blazar jets. 

\subsection{Flaring in a typical BL Lac} \label{BLsection}

BL Lac type blazars are typically less powerful, with slower bulk Lorentz factor jets than FSRQs. Their spectra tend to be dominated by jet emission with weak or absent disc emission. The $\gamma$-ray emission is usually well-fitted by SSC emission alone (\citealt{1998MNRAS.299..433F} and \citealt{2015MNRAS.453.4070P}). Markarian 501 is a typical high synchrotron peak frequency BL Lac in these regards and we have previously fitted its spectrum using our quiescent model \citep{2015MNRAS.453.4070P}. Transient particle acceleration events occurring at distances of $10r_{\m s}$, $10^{2}r_{\m s}$, $10^{3}r_{\m s}$ and $10^{4}r_{\m s}$ along the jet are shown in Fig. \ref{BLflare1}. 

\begin{figure*}
          \centering
          \includegraphics[width=17cm, clip=true, trim= 1cm 4.5cm 0cm 7cm]{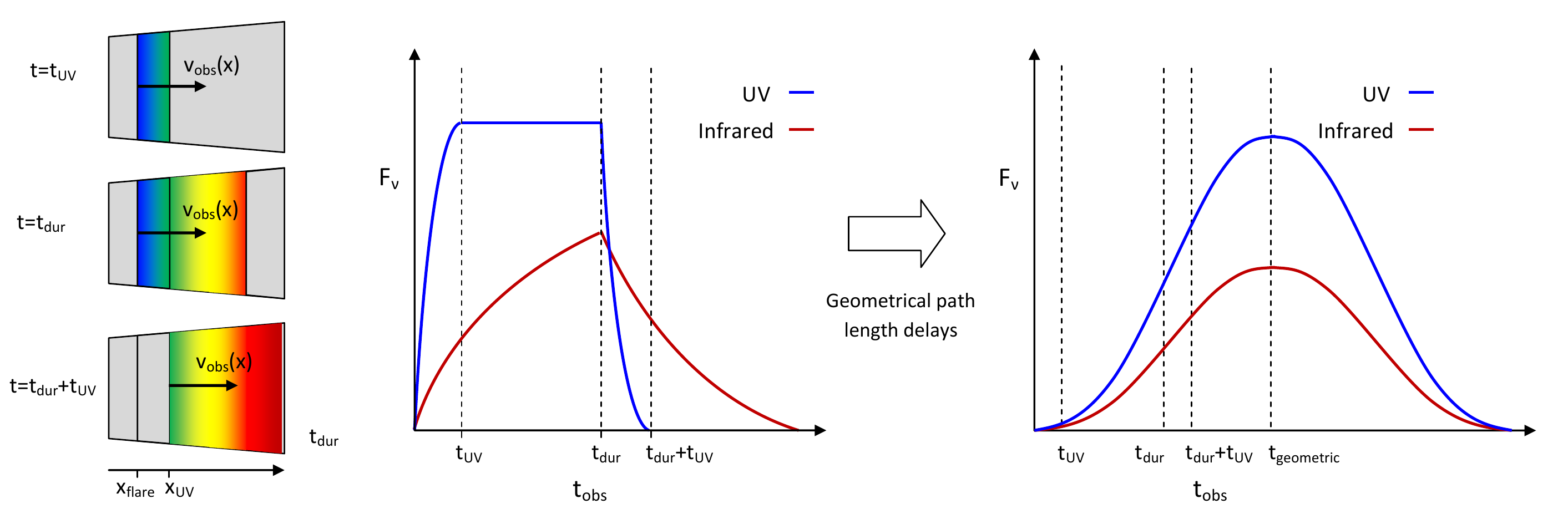}
          \caption{Schematic illustrating the lightcurve originating from a transient flaring event. At t=0, a flaring front at $x_{\m{flare}}$ starts accelerating non-thermal electrons in the plasma passing through the front. Colour is used to show the peak emission frequency of the plasma and has been chosen for convenience to correspond approximately to the visible spectrum (grey represents highly magnetised, dimly emitting plasma). Both the UV and IR emission of the flaring plasma initially increase with time because the volume of emitting plasma increases. In this example the radiative lifetime of UV emitting electrons, $t_{UV}$ is shorter than the flare duration $\Delta t_{\m{flare}}$. This means that for $t \gtrsim t_{UV}$ the UV emitting electrons which were initially accelerated at $t=0$ will have cooled and so the volume of UV emitting plasma will reach a maximum. This results in the plateau of the emitted UV flux in the lightcurve. The length of this maximum emitting plasma will be determined by the radiative lengthscale at UV wavelengths (the distance travelled during the radiative lifetime). This explains why the intrinsic rise time of the luminosity at a particular frequency is approximately given by the radiative cooling timescale of the electrons emitting at that frequency. The radiative lifetime of IR emitting electrons, however, is longer than the flaring duration in this example. The volume of IR emitting plasma increases throughout the flaring duration, since its rise time (or radiative lifetime) exceeds the flare duration. Once the particle acceleration at the flaring front switches off at $t=\Delta t_{\m{flare}}$ the flux at both frequencies decreases as the volume of existing emitting plasma shrinks due to radiative cooling. Because this process is effectively the time-reversal of the initial increase in volume of luminous plasma after the start of the flare, the rise and decay timescales are approximately equal and both are given by the radiative lifetime. It is important to note that at optically thick radio wavelengths lightcurves are complicated by radio lags and the non-negligable change in the jet radius over the radiative lengthscale, leading to longer decay timescales (see section \ref{radiosection} for details). In addition to the intrinsic rise and decay timescales which are given by the radiative lifetime, observed flares will also be smoothed due to geometrical path length differences between the observer and light traveling from different points along the flaring front. This acts to smooth the lightcurve on a timescale $t_{\m{geometric}}\sim R\sin(\theta_{\m{obs}})/c$ (see equation \ref{geom}), so the observed lightcurve will have rise and decay timescales equal to the longest out of the radiative and geometrical timescales. The effect of this geometrical delay is illustrated in the rightmost plot for the case where the geometrical delay is longer than the flaring duration. This results in the symmetric smoothing of the lightcurves on $t_{\m{geometric}}$. This effect is shown explicitly in figure \ref{comparedelay}. }
          \label{schematic3}
\end{figure*}

Figure \ref{BLflare1} shows that as the distance of the flaring location increases along the jet the luminosity of the flare decreases. This is because the magnetic field strength decreases and the radius of the flaring region increases further along the jet and so the synchrotron and SSC luminosity both decrease along the jet and the flare becomes less Compton-dominant. We find that both the rise and decay times of the flare at different wavelengths are approximately equal to the radiative lifetime of the electrons emitting at those wavelengths, smoothed on the geometrical delay timescale (\ref{geometrictimescale}). The intrinsic rise and decay timescales depend on the radiative lifetime because the length of luminous flaring plasma will be determined by the distance the plasma can travel past the flaring front during the radiative lifetime of electrons emitting at a given frequency. Once a flaring particle acceleration event begins, the luminosity will grow as the length of luminous plasma following the flaring front increases. The luminosity of the flare will then plateau once the duration of the flaring event exceeds both the radiative lifetime and the effective length of flaring plasma reaches a maximum (the radiative lengthscale). Once the flaring event stops, particle acceleration at the flare ceases and no new flaring plasma will be created. The flare will then decay over the radiative lifetime at the observed frequency with a time-dependence which is approximately the time reversal of the flare rise. This intrinsic rise and decay will then be smoothed on the geometrical delay timescale. The effect of the radiative lifetime and geometric delay on the flaring lightcurve is illustrated in Figures \ref{schematic3} and \ref{comparedelay}. 

The spectral evolution of a flaring event with a constant power injected into accelerating electrons, $\epsilon_{\m{flare}}=1$, at the flaring front is shown in Figure \ref{BLflare1}. At frequencies for which both the radiative lifetime and gemoetrical delay timescale are shorter than the flare duration we observe a flat plateau in luminosity. If instead, the power injected into particle acceleration varied smoothly with time, the behaviour would become more complicated. The effective response function of the plasma to changes in the power injected by particle acceleration at the flare is given by the longer timescale out of the radiative lifetime and geometrical delay (as shown in Fig. \ref{comparedelay}). This means that the lightcurve will be smoothed on both the radiative lifetime and geometrical delay timescale at a particular frequency. Variations in flaring power which are shorter than either timescale at a given frequency will be smoothed out and not easily visible. However, variations on timescales exceeding both the radiative lifetime and geometric delay timescale will be clearly visible in the lightcurve (since the variations will be slower than the intrinsic response time of the flaring plasma to variations in the flaring properties). 

This leads to the first prediction of this work - observed flaring lightcurves should come in two main types:
\begin{itemize}
\item {\bf 1 - Symmetric flares} occur on timescales shorter than either the radiative lifetime or geometric delay timescale at the observed frequency. Their rise and decay timescales will be approximately equal and given by the longest timescale out of the radiative lifetime of electrons emitting at that frequency and the geometrical delay (\ref{geometrictimescale}).\footnote{At radio frequencies the radiative lifetime may be so long that the plasma is able to travel large enough distances along the jet that the jet radius increases substantially. In this case the flare may no longer appear symmetric since the decay time can be longer than the rise time of the radio flare. This is because the radiative lifetime increases along the jet as the magnetic field strength decreases (see section \ref{radiosection} for more discussion).     }
\item {\bf 2 - Extended flares} occur when the duration of the flaring event exceeds both the radiative lifetime at the observed frequency and the geometric delay timescale. Their lightcurves will show structure which tracks the time-dependent power injected into particle acceleration by the flaring event (smoothed on both the radiative lifetime and geometrical delay timescale). 
\end{itemize}

Symmetric flares are useful for determining the approximate radiative lifetime of the emitting electrons. Extended flares tell us about the physics responsible for the flaring event itself and how the power of the particle acceleration process changes with time. 

\subsection{A Physical Explanation for Orphan flares}

\begin{figure*}
	\centering
		\subfloat[$x_{\m{flare}}=4.02\times10^{-3}\m{pc}$,\qquad $\Delta t_{\m{flare}}=$8.63 hours]{ \includegraphics[width=6.8cm, clip=true, trim=0cm 0cm .cm 0cm]{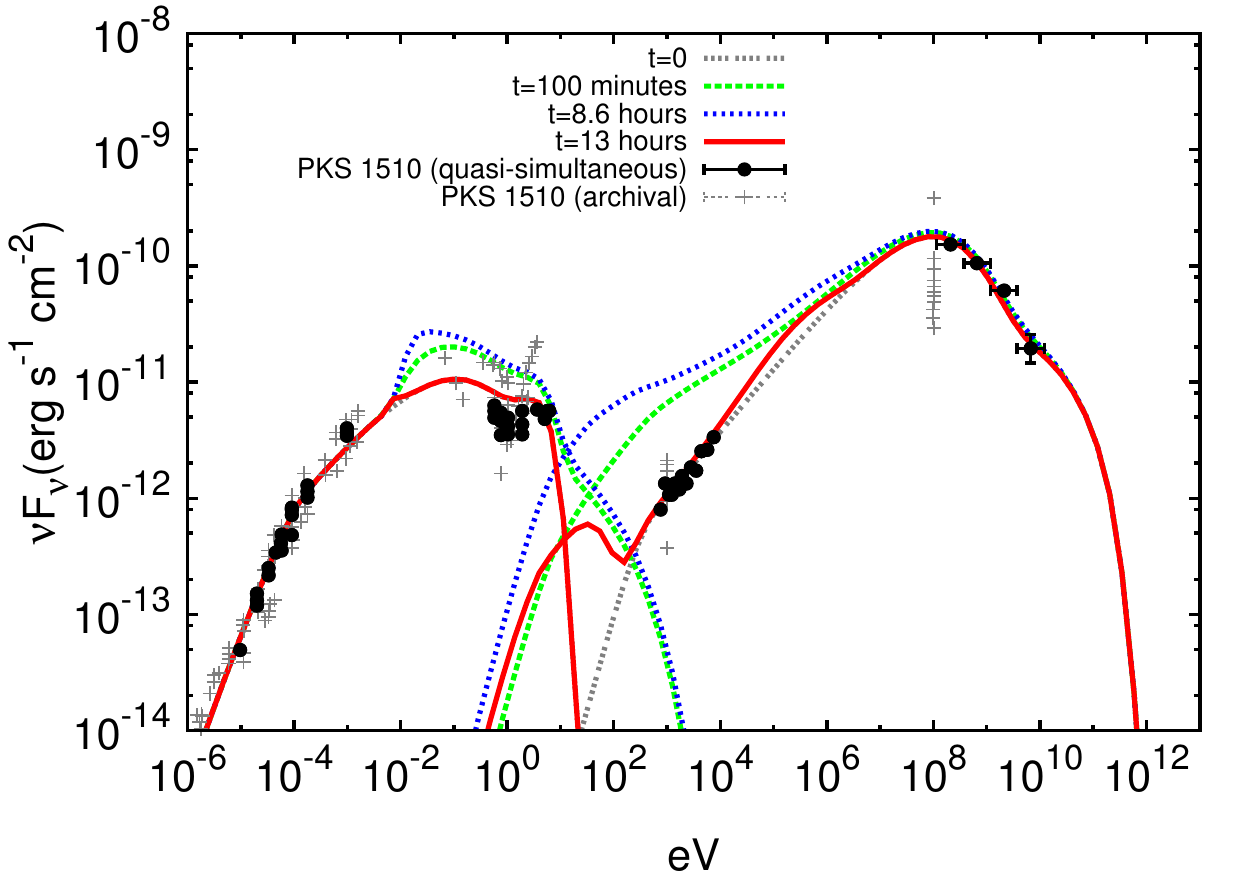} }
\qquad
		\subfloat[$x_{\m{flare}}=4.02\times10^{-3}\m{pc}$,\qquad $\Delta t_{\m{flare}}=$8.63 hours]{ \includegraphics[width=6.8cm, clip=true, trim=0cm 0cm 0.cm 0cm]{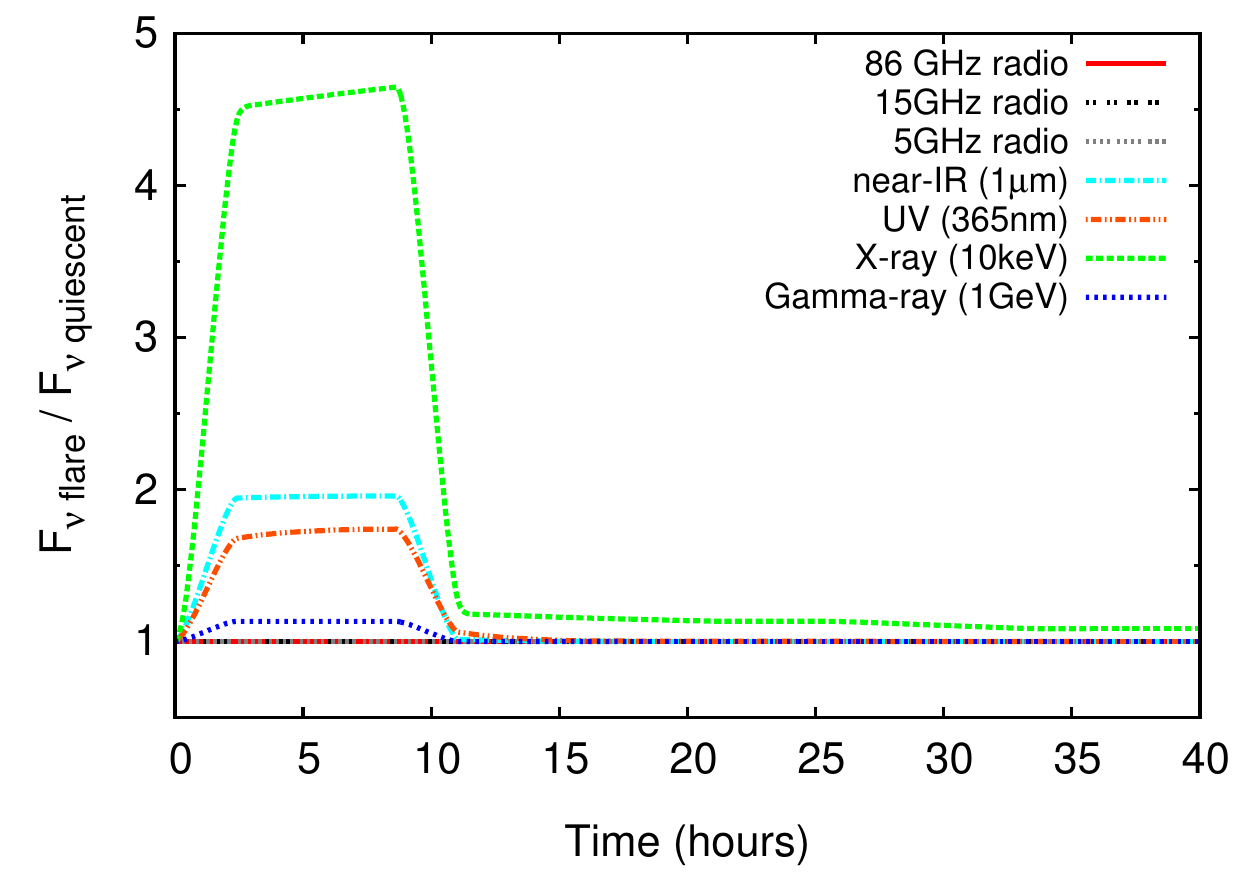} }
\vspace{-0.2cm}
		\subfloat[$x_{\m{flare}}=0.0402\m{pc}$,\qquad $\Delta t_{\m{flare}}=$21.0 hours]{ \includegraphics[width=6.8cm, clip=true, trim=0cm 0cm 0.cm 0cm]{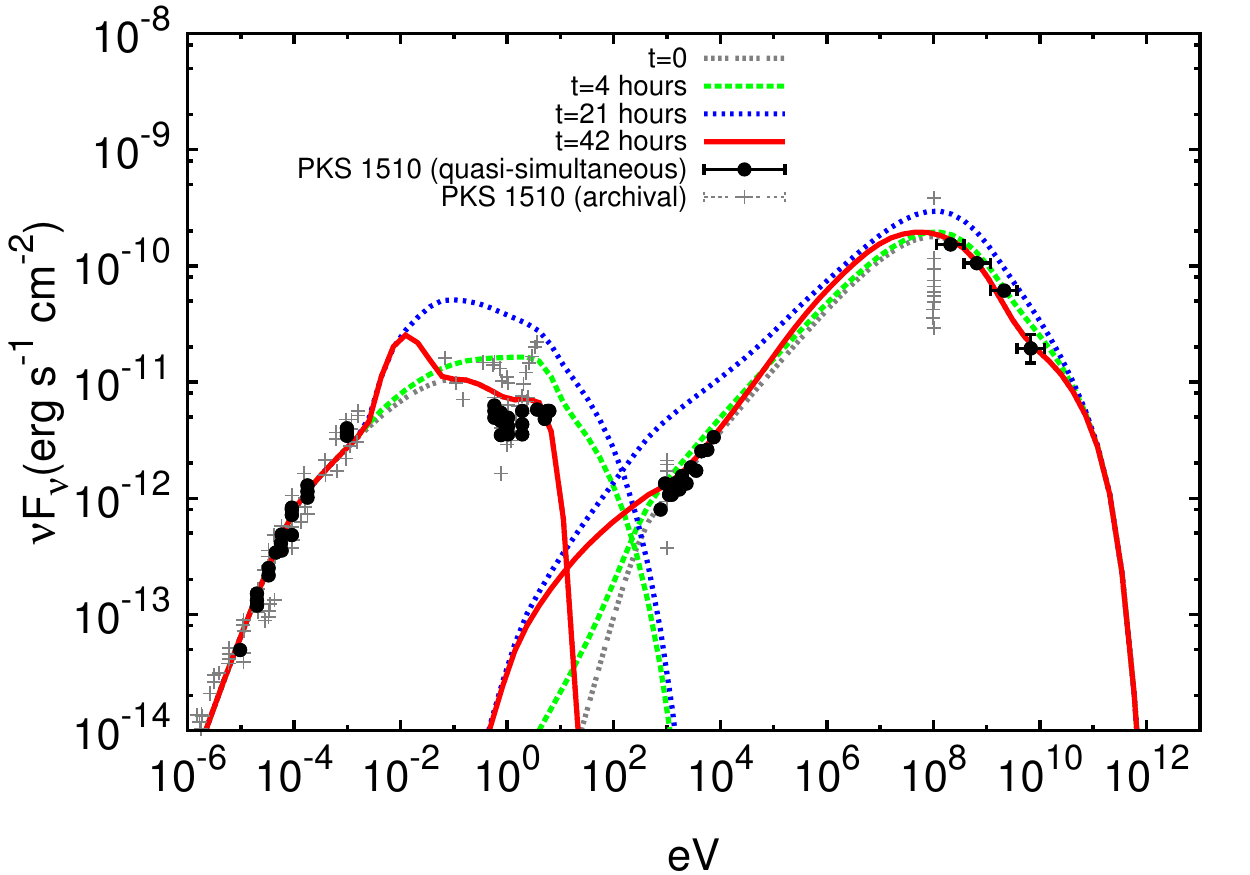} }
\qquad
		\subfloat[$x_{\m{flare}}=0.0402\m{pc}$,\qquad $\Delta t_{\m{flare}}=$21.0 hours]{ \includegraphics[width=6.8cm, clip=true, trim=0cm 0cm 0.cm 0cm]{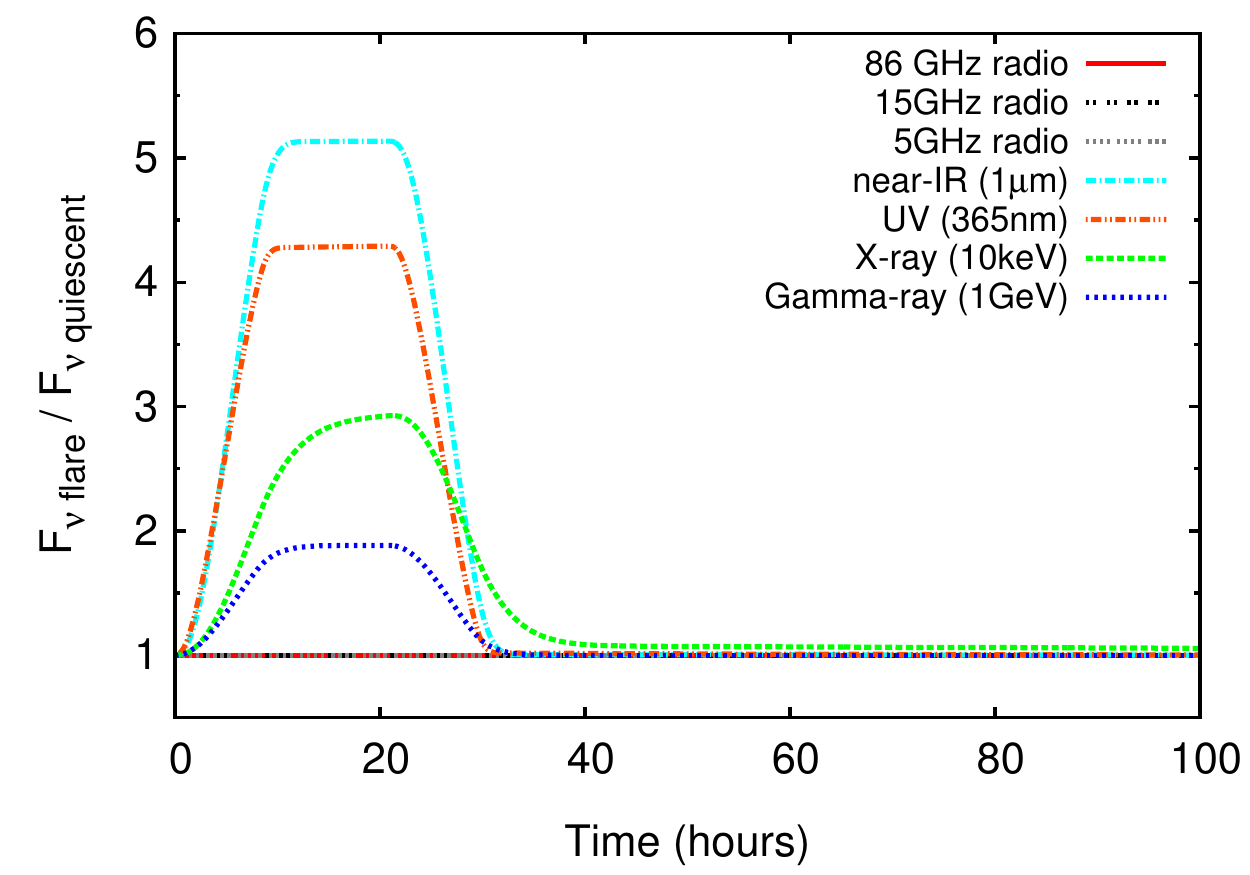} }
\vspace{-0.2cm}
\subfloat[$x_{\m{flare}}=0.402\m{pc}$,\qquad $\Delta t_{\m{flare}}=$2.06 days]{ \includegraphics[width=6.8cm, clip=true, trim=0cm 0cm 0.cm 0cm]{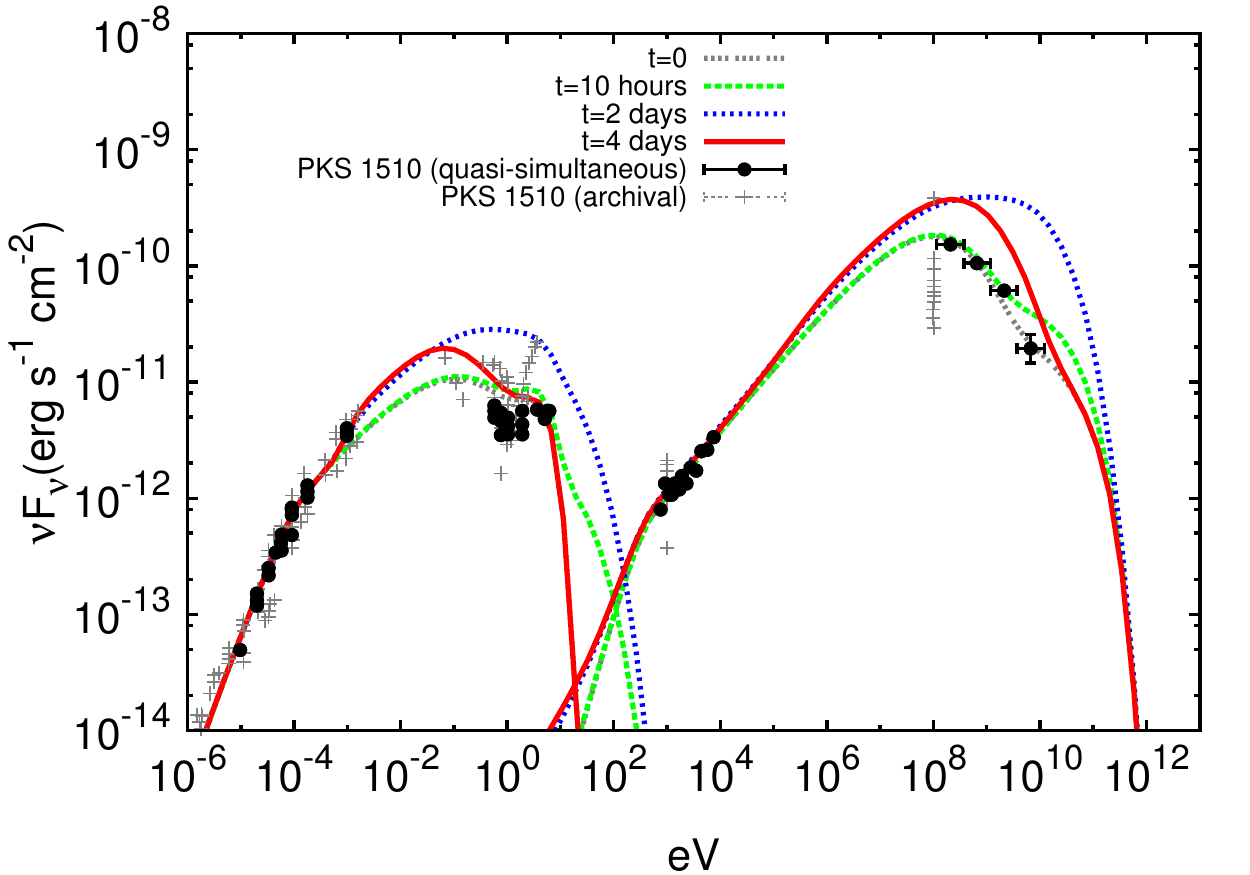} }
\qquad
		\subfloat[$x_{\m{flare}}=0.402\m{pc}$,\qquad $\Delta t_{\m{flare}}=$2.06 days]{ \includegraphics[width=6.8cm, clip=true, trim=0cm 0cm 0.cm 0cm]{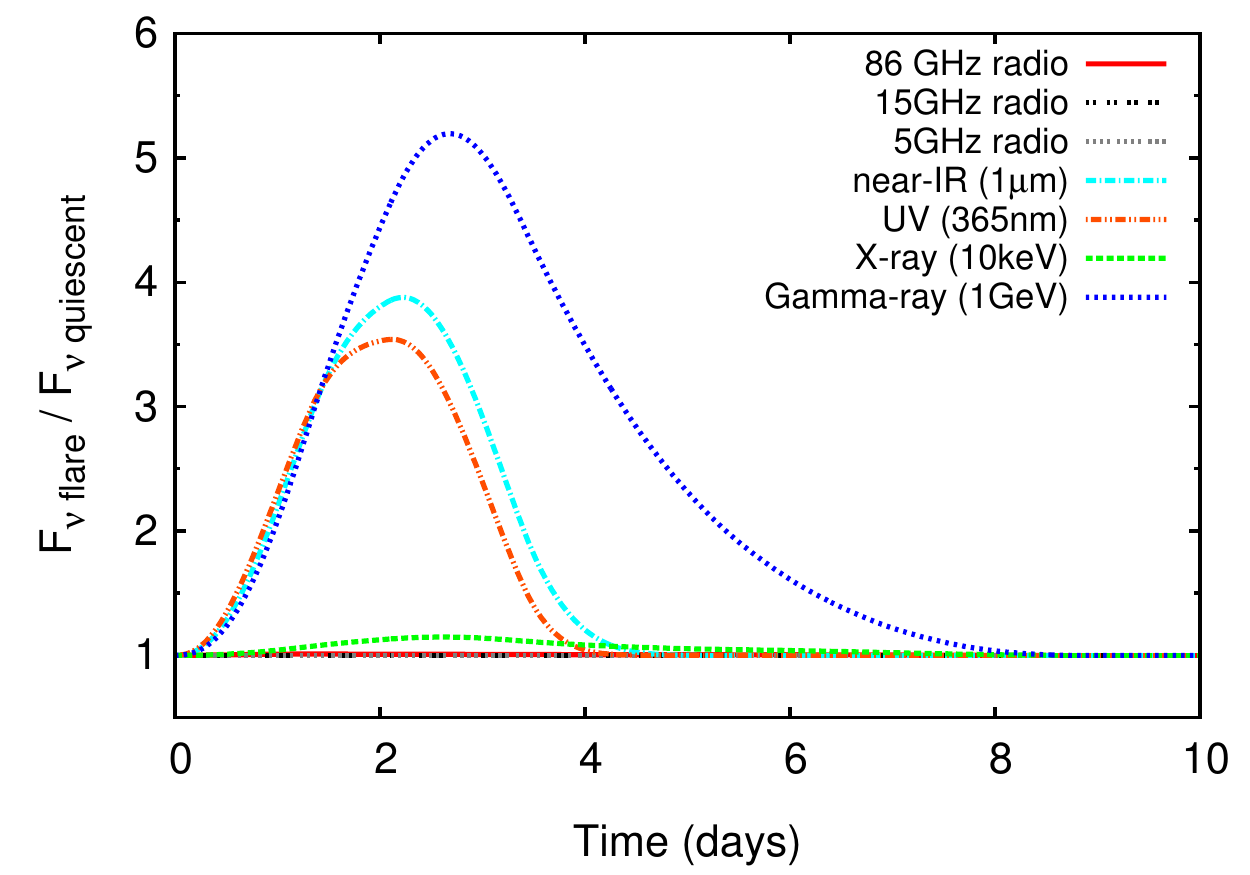} }
\vspace{-0.2cm}
\subfloat[$x_{\m{flare}}=4.02\m{pc}$,\qquad $\Delta t_{\m{flare}}=$4.93 days]{ \includegraphics[width=6.8cm, clip=true, trim=0cm 0cm 0.cm 0cm]{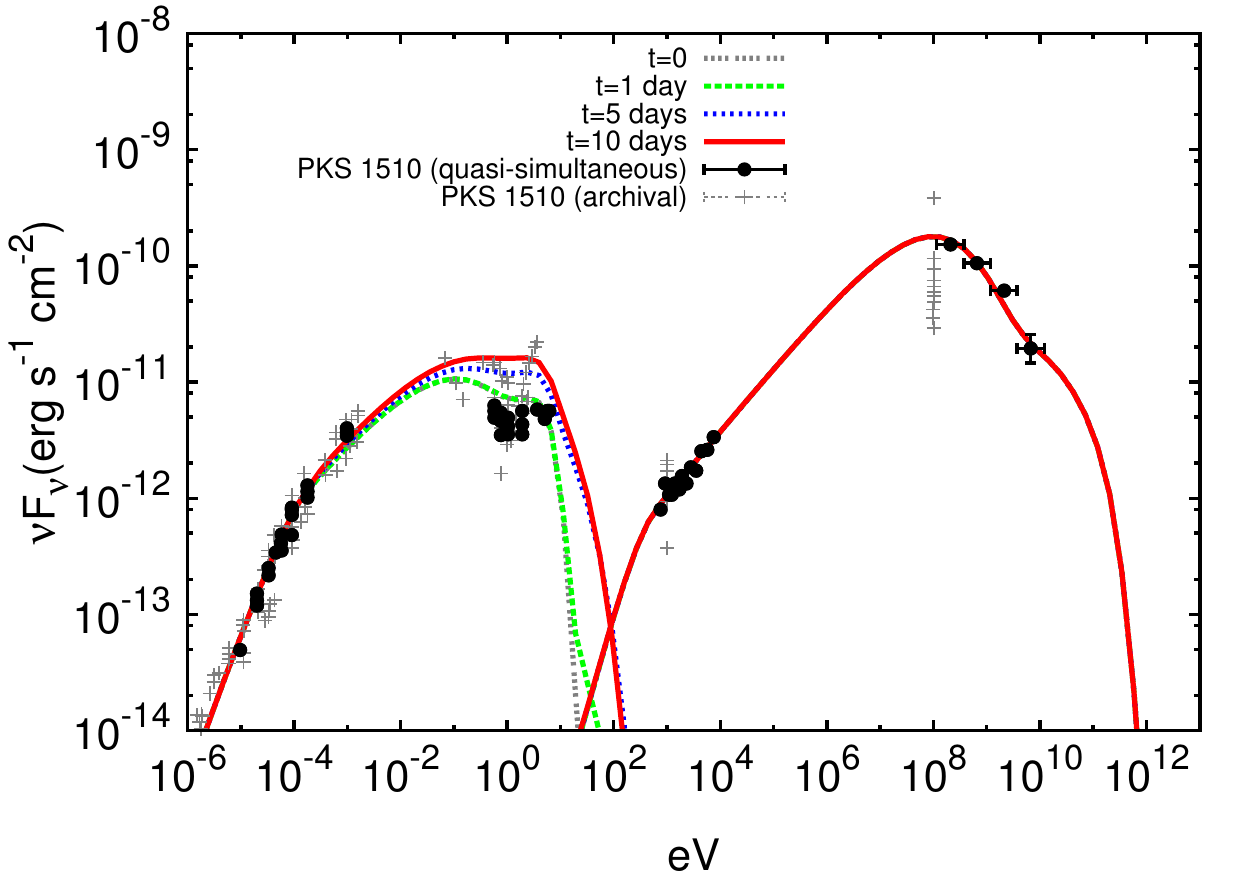} }
\qquad
		\subfloat[$x_{\m{flare}}=4.02\m{pc}$,\qquad $\Delta t_{\m{flare}}=$4.93 days]{ \includegraphics[width=6.8cm, clip=true, trim=0cm 0cm 0.cm 0cm]{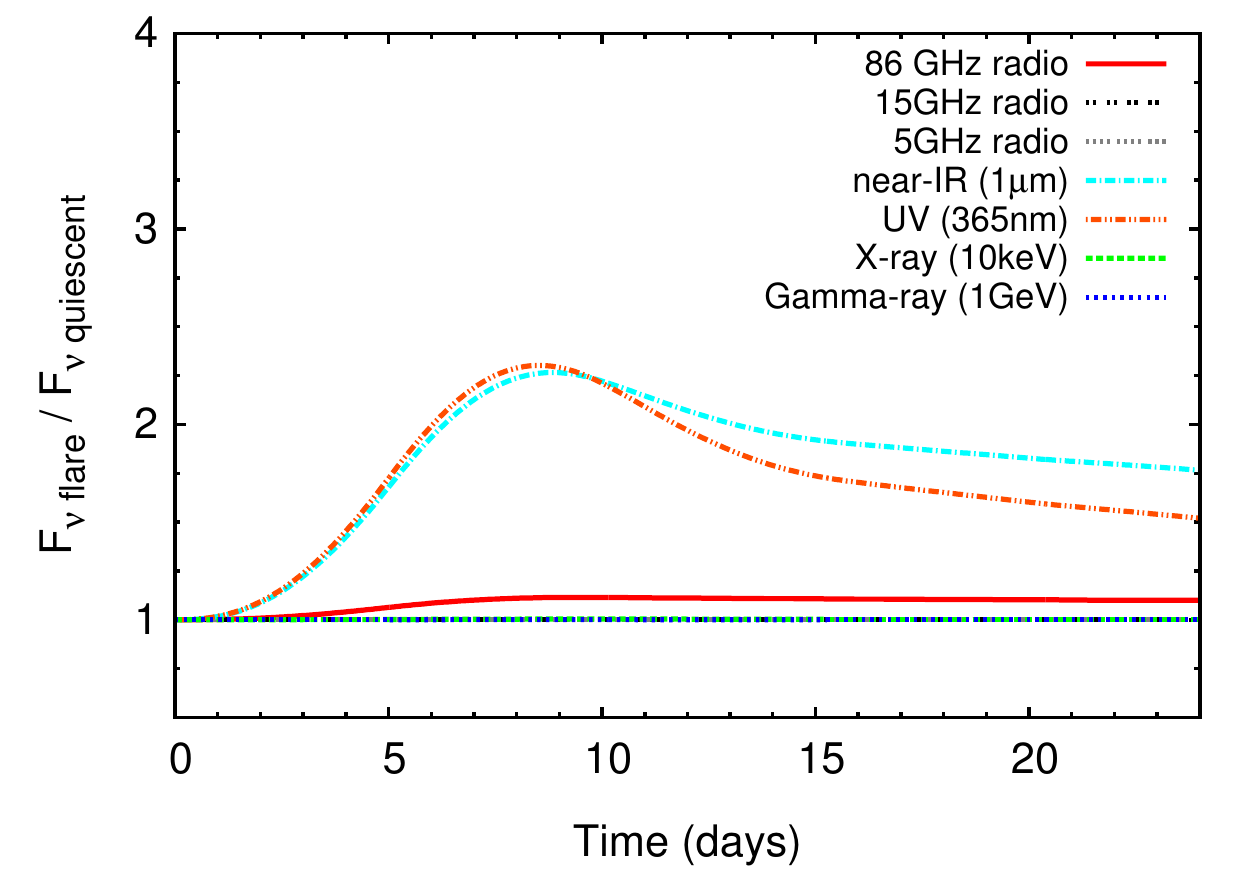} }	
		
	\caption{The evolution of the spectrum and lightcurves caused by flaring events at different locations, $x_{\m{flare}}$, along the parabolic base of the jet in PKS1510-089 (the transition region occurs at $40.2\m{pc}$). The flaring luminosity and Compton-dominance are a complicated function of the distance of the flaring front along the jet $x_{\m{flare}}$. Close to the jet base the magnetic field strength is largest and the jet radius smallest. This results in an increase in X-ray SSC emission and emission around the synchrotron peak frequency (a). At larger distances along the parabolic base the combined synchrotron and SSC flaring luminosity decreases, whilst the flare becomes less Compton-dominant because the jet radius increases and the magnetic field strength decreases along the jet: (c) and (g). Flares which occur close to the outer edge of the BLR can produce Compton-dominant high energy $\gamma$-ray flares by inverse-Compton scattering BLR photons, this is because here the energy density of external BLR photons can be much larger than the energy density of synchrotron photons.  } 
\label{J0531flare1}
\end{figure*}

\begin{figure*}
	\centering

		\subfloat[Same as (b) but without geometrical time delay.]{ \includegraphics[width=6.8cm, clip=true, trim=0cm 0cm 0.cm 0cm]{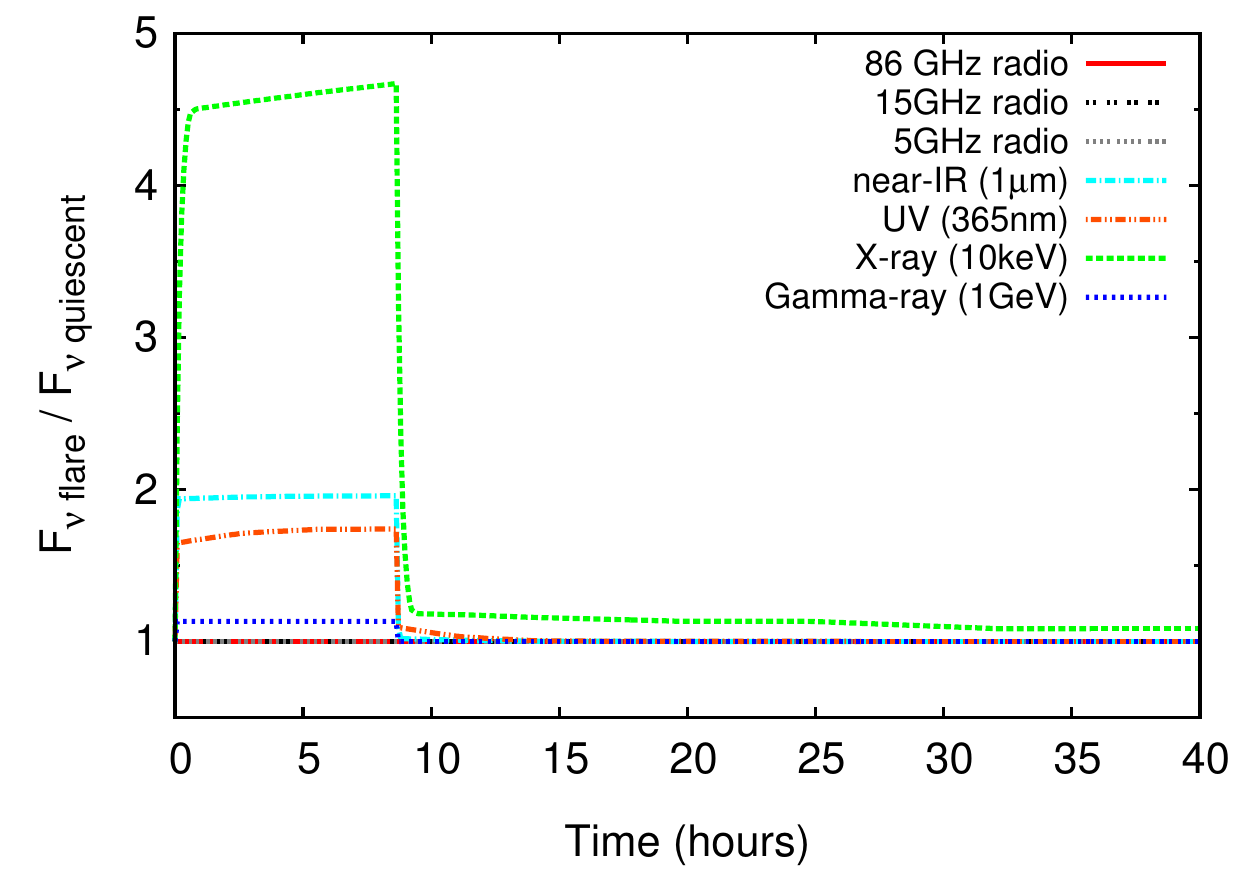} }
\qquad
		\subfloat[$x_{\m{flare}}=4.02\times10^{-3}\m{pc}$,\qquad $\Delta t_{\m{flare}}=$8.63 hours]{ \includegraphics[width=6.8cm, clip=true, trim=0cm 0cm 0.cm 0cm]{PKS1510lightcurve_smeared_r=10.pdf} }
\vspace{-0.2cm}
		\subfloat[Same as (d) but without geometrical time delay.]{ \includegraphics[width=6.8cm, clip=true, trim=0cm 0cm 0.cm 0cm]{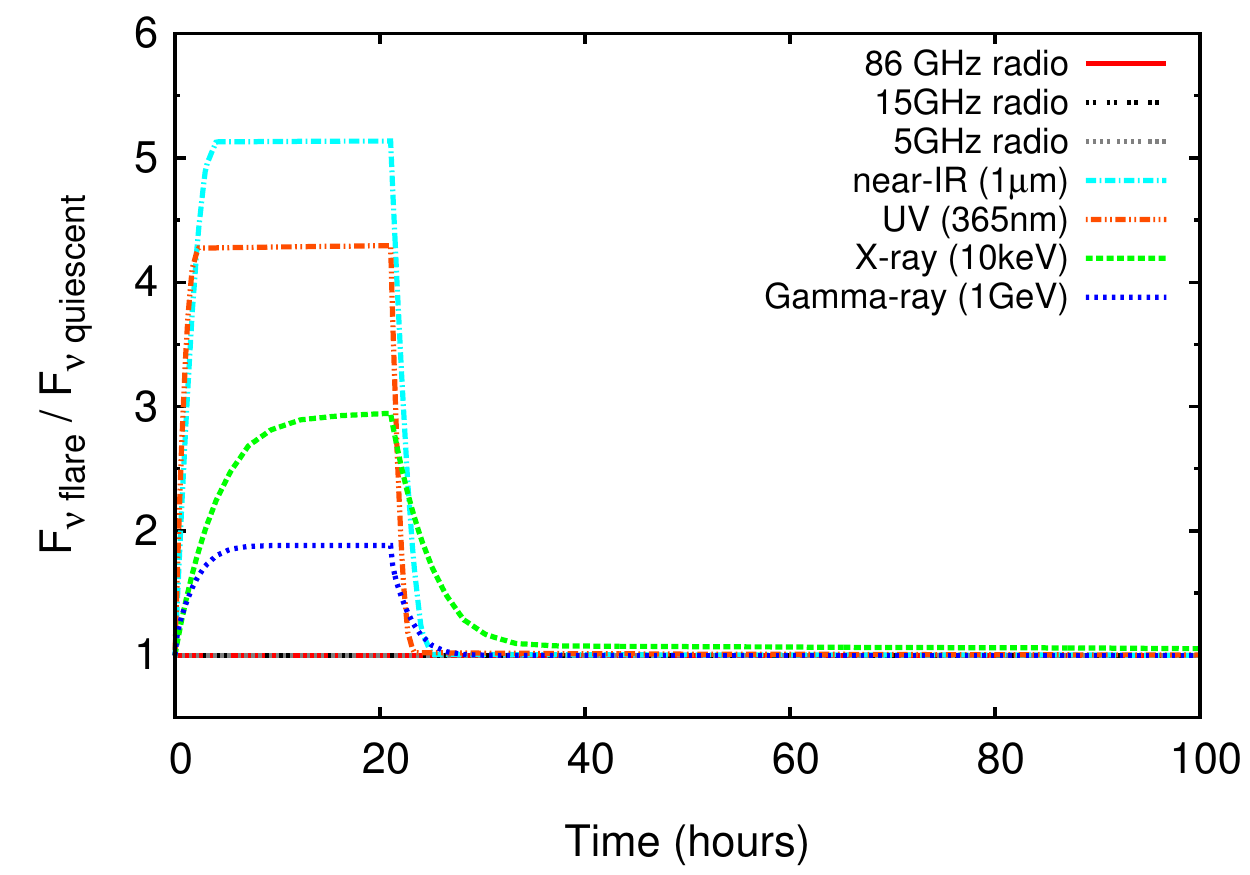} }
\qquad
		\subfloat[$x_{\m{flare}}=0.0402\m{pc}$,\qquad $\Delta t_{\m{flare}}=$21.0 hours]{ \includegraphics[width=6.8cm, clip=true, trim=0cm 0cm 0.cm 0cm]{PKS1510lightcurve_smeared_r=10_2.pdf} }
\vspace{-0.2cm}
		\subfloat[Same as (f) but without geometrical time delay.]{ \includegraphics[width=6.8cm, clip=true, trim=0cm 0cm 0.cm 0cm]{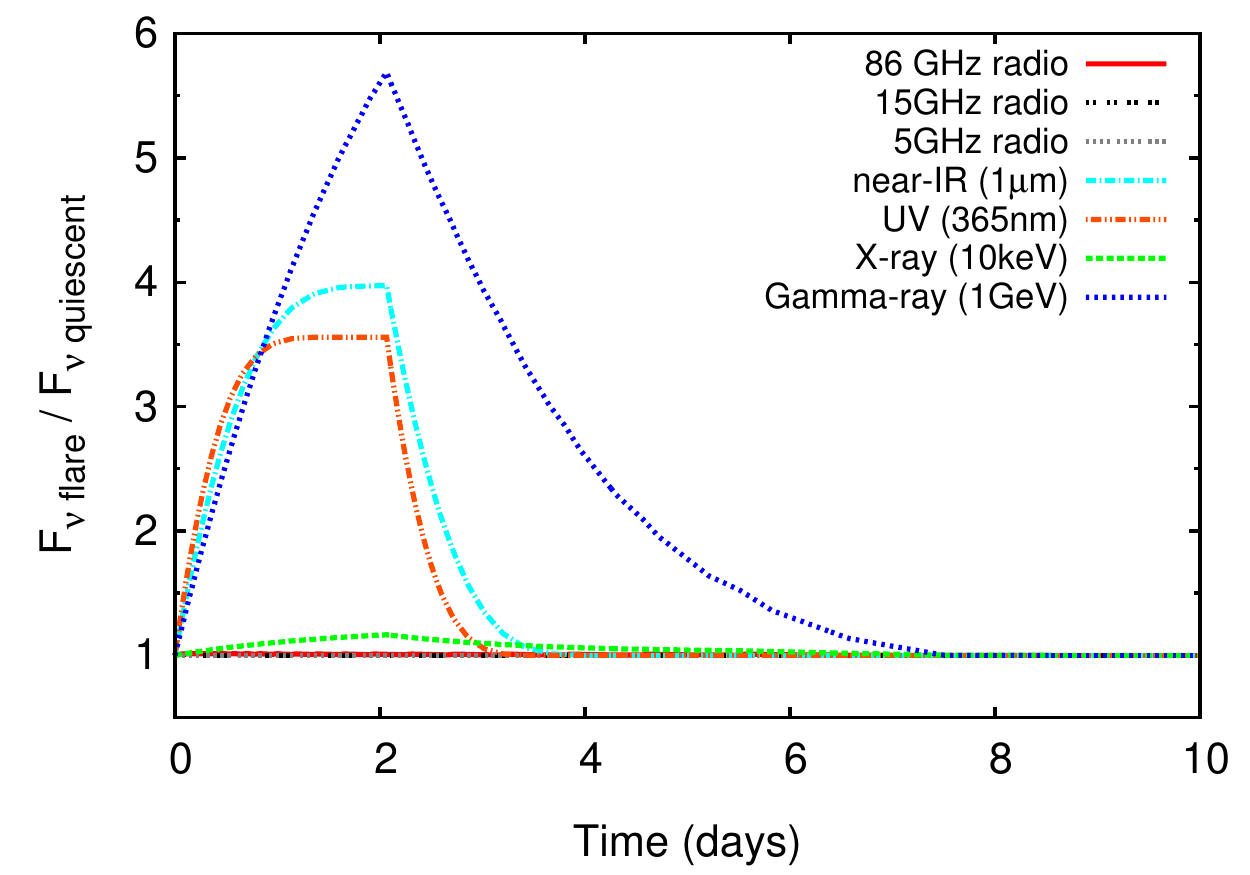} }
\qquad
		\subfloat[$x_{\m{flare}}=0.402\m{pc}$,\qquad $\Delta t_{\m{flare}}=$2.06 days]{ \includegraphics[width=6.8cm, clip=true, trim=0cm 0cm 0.cm 0cm]{PKS1510lightcurve_smeared_r=10_3.pdf} }
\vspace{-0.2cm}
		\subfloat[Same as (h) but without geometrical time delay.]{ \includegraphics[width=6.8cm, clip=true, trim=0cm 0cm 0.cm 0cm]{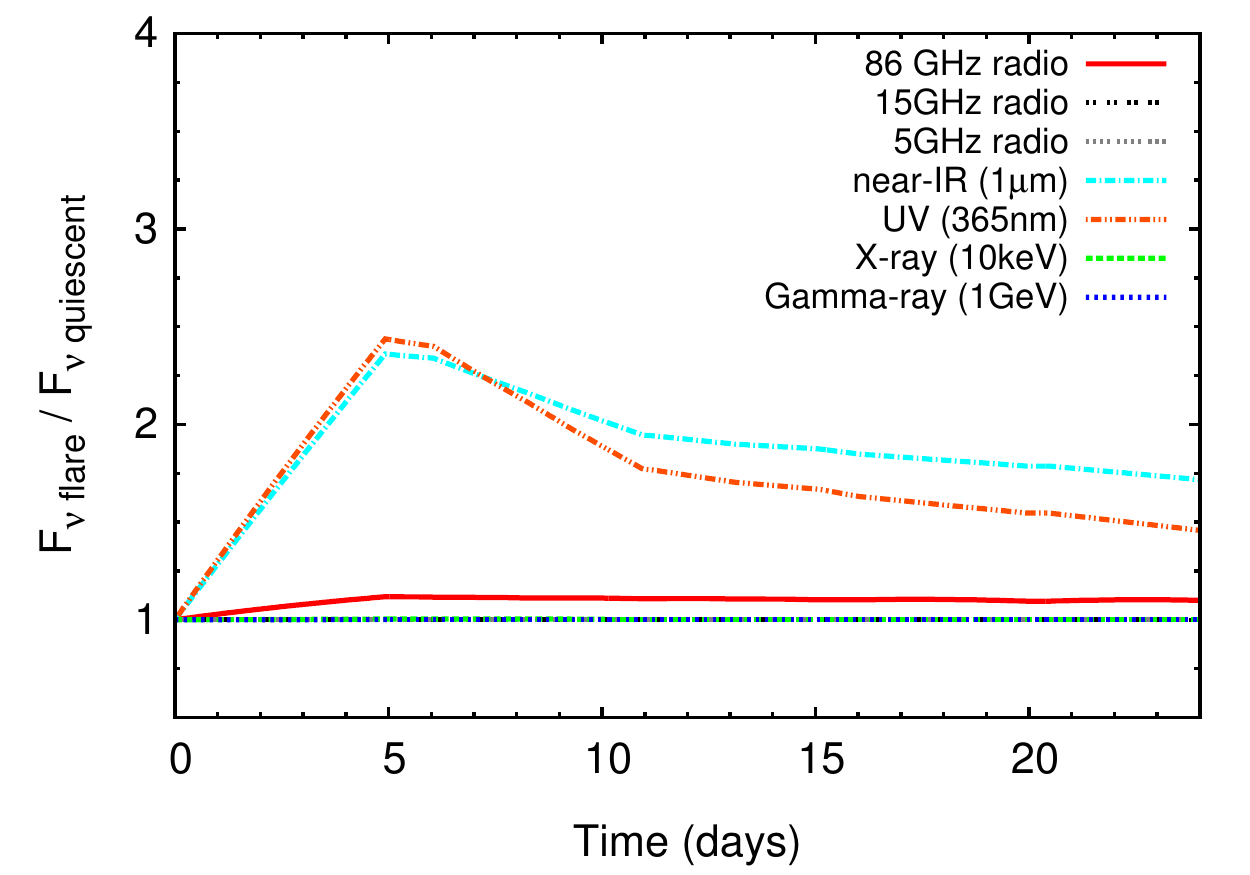} }
\qquad
		\subfloat[$x_{\m{flare}}=4.02\m{pc}$,\qquad $\Delta t_{\m{flare}}=$4.93 days]{ \includegraphics[width=6.8cm, clip=true, trim=0cm 0cm 0.cm 0cm]{PKS1510lightcurve_smeared_r=10_4.pdf} }	
		
	\caption{Figure illustrating the PKS1510 flaring lightcurves calculated with and without the inclusion of geometrical time delay which is caused by differences in the path length between the observer and different points on the flaring front (see section \ref{geometricaldelay}). In the left column are the same flaring lightcurves for PKS1510 shown in fig. \ref{J0531flare1}, but without inclusion of the geometrical time delay (\ref{geom}). For comparison, the right column shows the same flaring lightcurves including the geometrical delay. The figure shows that when the geometrical time delay is smaller than the radiative lifetime the rise and decay of the flare occurs on a radiative lifetime (e.g. the X-ray lightcurve in fig. d) whilst in the case that the geometrical time delay is longer than the radiative lifetime the lightcurve is instead smoothed on the geometrical timescale (e.g. the near-IR lightcurve in fig. b). For a detailed explanation see the schematic fig. \ref{schematic3}.   } 
\label{comparedelay}
\end{figure*}

A simple consequence of the rise and decay timescale of a flaring event being approximately the radiative lifetime at each observed frequency, is that the flaring luminosity of the highest energy synchrotron and inverse-Compton emission will rise fastest. This is because the synchrotron and inverse-Compton power emitted by electrons increases rapidly with electron energy and the frequency of the synchrotron and inverse-Compton radiation also increases with electron energy. This means that the electrons responsible for the highest frequency synchrotron and inverse-Compton emission will have the shortest radiative lifetimes. Once a flare begins, the highest frequency synchrotron and inverse-Compton emission will reach their peak luminosity first, followed by progressively lower frequencies (provided we are not in the Klein-Nishina regime). Flares which are short in duration will therefore only be visible at the highest synchrotron and inverse-Compton frequencies. This is because these are the only frequencies for which the radiative lifetime of the emitting electrons is shorter than the flare duration and will reach their maximum luminosity, becoming luminous enough to be visible above the quiescent jet emission. We therefore expect that short duration flares should only be visible at $\gamma$-ray and/or the peak synchrotron frequency (X-ray/UV/optical). Which combination of the flaring synchrotron and inverse-Compton peaks will be observed depends on the luminosity and Compton-dominance of the flaring and quiescent emission, since the flaring emission must be observable above the quiescent emission. A Compton-dominant flare which is sufficiently luminous to be observed above the quiescent emission is likely to be classified as an inverse-Compton orphan flare (observed at $\gamma$-ray energies), whereas a luminous synchrotron dominant flare is likely to be classified as a synchrotron orphan flare (observed at optical/UV/X-ray frequencies depending on the synchrotron peak frequency). This provides a physical explanation of the puzzling \lq{}orphan flare\rq{} phenomenon which has been of much interest recently: short flares visible only at either high energy $\gamma$-ray or X-ray wavelengths with no corresponding increase in luminosity at other frequencies (e.g. \citealt{2004ApJ...601..151K} and \citealt{2013A&A...552A..11R}). Orphan inverse-Compton and synchrotron flares are naturally produced in our leptonic model as shown in Figures \ref{BLflare1} and \ref{J0531flare1}.  

It is important to emphasise that it is only because we model the flaring event in addition to the quiescent emission that we are able to understand and reproduce orphan flares. In a one-zone model all the observed emission comes from the flaring region and so all frequencies will be observed to rise in luminosity during a flaring event. It is no surprise therefore, that it is difficult to understand orphan flares in the context of one-zone models (\citealt{2005ApJ...630..130B} and \citealt{2013A&A...552A..11R}). In our model the emission from the flare has to be sufficiently luminous to be observed above the quiescent emission. The flaring luminosity at low synchrotron and inverse-Compton frequencies increases on very long timescales and so it is not able to increase in luminosity sufficiently to exceed the quiescent luminosity in the duration of a short flaring event, leading to an orphan flare. In the flares calculated in this work the flaring plasma has been chosen to be in, or close to, equipartition. However, it is important to note that the equipartition fraction or magnetisation of a flare also affects its Compton-dominance, with particle-dominated flares tending to be more Compton-dominant than magnetically dominated flaring plasma. Thus the magnetisation of the flaring plasma also has an important effect on whether an observed flare will be classified as an orphan synchrotron or orphan inverse-Compton flaring event.  

\subsection{Flaring in a typical FSRQ}

Let us now consider flaring in a typical FSRQ type blazar PKS1510-089. The main spectral differences between PKS1510 and Mkn501 are that PKS1510 has a higher Compton-dominance and lower peak emission frequencies. From modelling, it is generally found that FSRQs have larger jet powers than BL Lacs and the inverse-Compton $\gamma$-ray emission in FSRQs is dominated by scattering external photons instead of being dominated by scattering synchrotron photons, as in BL Lacs. These differences are typical of the broad spectral differences between BL Lacs and FSRQs \citep{2015MNRAS.453.4070P}. The quiescent emission of PKS1510 has been fitted using model parameters in Table \ref{Table2}, where crucially the majority of the $\gamma$-ray emission originates from inverse-Compton scattering of CMB and starlight seed photons. This is because the transition region is constrained to be at a large distance of 40.2pc by the optically thick to thin synchrotron break (see section 5 and Fig. 2 of \citealt{2013MNRAS.431.1840P}) and is well outside the expected radius of the BLR or dusty torus (\citealt{2005ApJ...629...61K}, \citealt{2006NewAR..50..728E} and \citealt{2008ApJ...685..160N}) .

The spectra calculated for flaring events occurring at distances of $10r_{\m s}$, $10^{2}r_{\m s}$, $10^{3}r_{\m s}$ and $10^{4}r_{\m s}$ along the jet are shown in Fig. \ref{J0531flare1}. As in section \ref{BLsection} we consider flaring events which are in causal contact and flaring particle acceleration which produces plasma in equipartition immediately upon passing through the flaring front (i.e. $\Delta_{\m{flare}}=1$ and $\epsilon_{\m{flare}}=1$). The lightcurves again show a characteristic rise and decay timescale at a given frequency equal to the radiative lifetime of the emitting electrons and smoothed by the observed geometrical time delay. The flaring SEDs have some similarities to those of Markarian 501. Flares at short distances along the jet cause a substantial increase in the SSC luminosity because of the high magnetic field strength and smaller jet radius close to the jet base. This is most clearly visible above the quiescent inverse-Compton emission at X-ray energies. At larger distances along the parabolic base, the magnetic field decreases and jet radius increases, so the flare becomes less Compton-dominant and is most clearly visible at peak synchrotron frequencies (optical/UV). The spectrum of the flaring region is essentially similar to a typical BL Lac spectrum. This is because the higher synchrotron luminosity and smaller jet radius close to the jet base mean that the dominant inverse-Compton seed photon source is now synchrotron photons instead of external seed photons. This results in the flaring spectrum becoming far less Compton-dominant than a typical quiescent FSRQ spectrum. This is the case for most flares occurring in the parabolic base, however, flares which happen to occur close to the outer edge of the BLR or dusty torus produce flaring spectra which are similar to a typical Compton-dominant FSRQ spectrum (for a flare occurring close to the outer edge of the BLR see Fig \ref{J0531flare1}e and \ref{fit}). This is because at distances close to the outer edge of the BLR or dusty torus, inverse-Compton scattering of external BLR or dusty torus seed photons will dominate over synchrotron and SSC emission.

These results show that the location of a flaring particle acceleration event leaves a distinct signature on both the observed flaring spectrum and lightcurve. The location of the flare determines the radiative lifetimes of the electrons emitting at different frequencies, along with the geometrical time delay. The radiative lifetime and geometrical delay timescale directly correspond to the observed rise and decay timescale of the flaring lightcurves at those frequencies. In Figure \ref{comparedelay} we explicitly show the flaring lightcurves for PKS1510 with and without the inclusion of the geometric delay (which is caused by path length differences between the observer and different points along the flaring front) to illustrate the separate effects of the radiative lifetime and geometrical time delay in determining the observed lightcurve. Close to the base of the jet, the high magnetic field strength and small jet radius mean that flares predominantly produce X-ray SSC emission and emission around the peak synchrotron frequency. At larger distances along the parabolic base, flares are observable predominantly at the synchrotron peak frequency. Compton-dominant flares are possible in FSRQs if the flaring location is close to the outer radius of the BLR or dusty torus. These results show that the location of a flare can be determined from observations of its multiwavelength spectrum and lightcurve (this will be demonstrated in section \ref{section6}). In the next section we shall deal with an alternative scenario in which flaring results from a transient change in the jet power, with the location of particle acceleration the same as in the quiescent jet (i.e. initially coming into equipartition at the transition region with additional in situ acceleration occurring along the conical section of the jet).

\begin{figure*}
	\centering
		\subfloat[$x_{\m{flare}}=0.304\m{pc}$, \qquad$\Delta t_{\m{flare}}=$10 days]{ \includegraphics[width=6.8cm, clip=true, trim=0cm 0cm 0.cm 0cm]{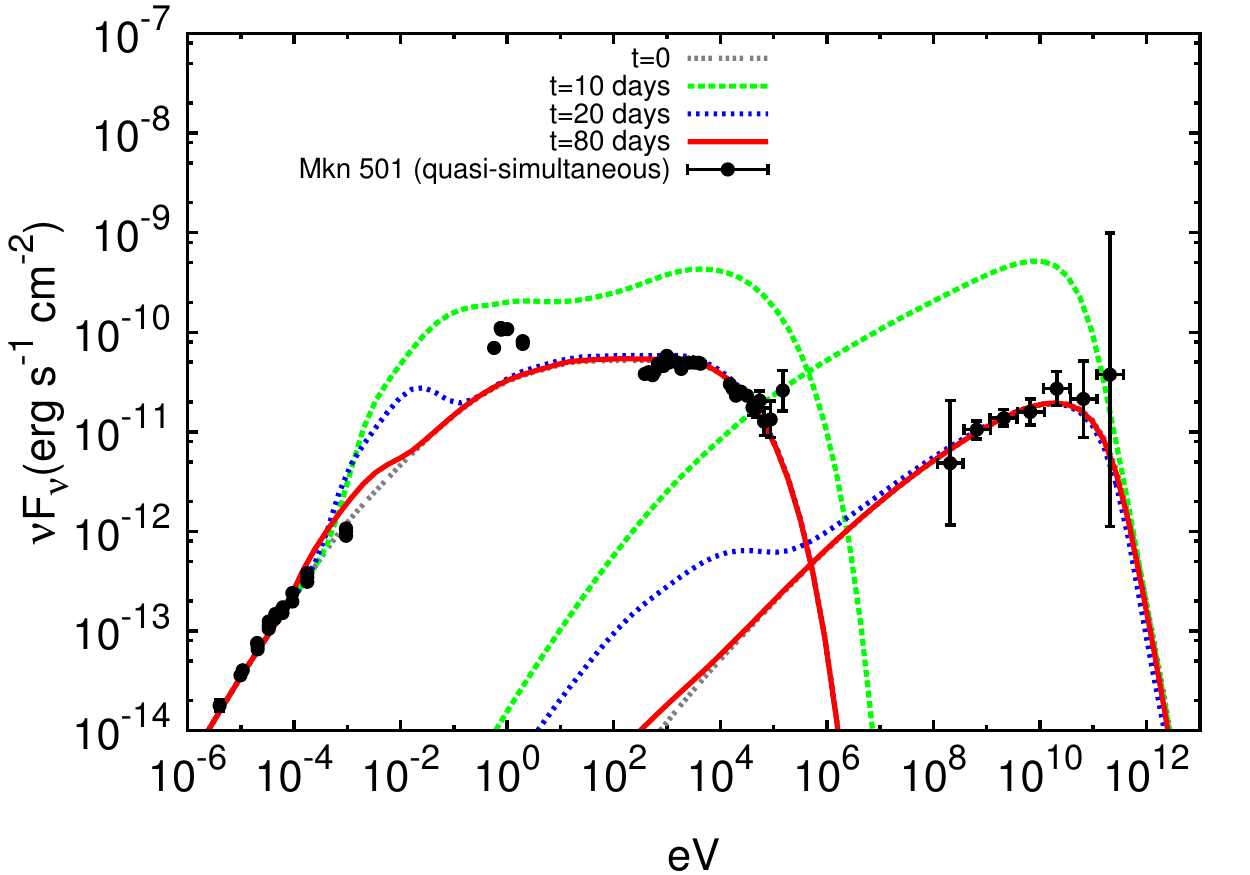} }
\qquad
		\subfloat[$x_{\m{flare}}=0.304\m{pc}$,\qquad $\Delta t_{\m{flare}}=$10 days]{ \includegraphics[width=6.8cm, clip=true, trim=0cm 0cm 0.cm 0cm]{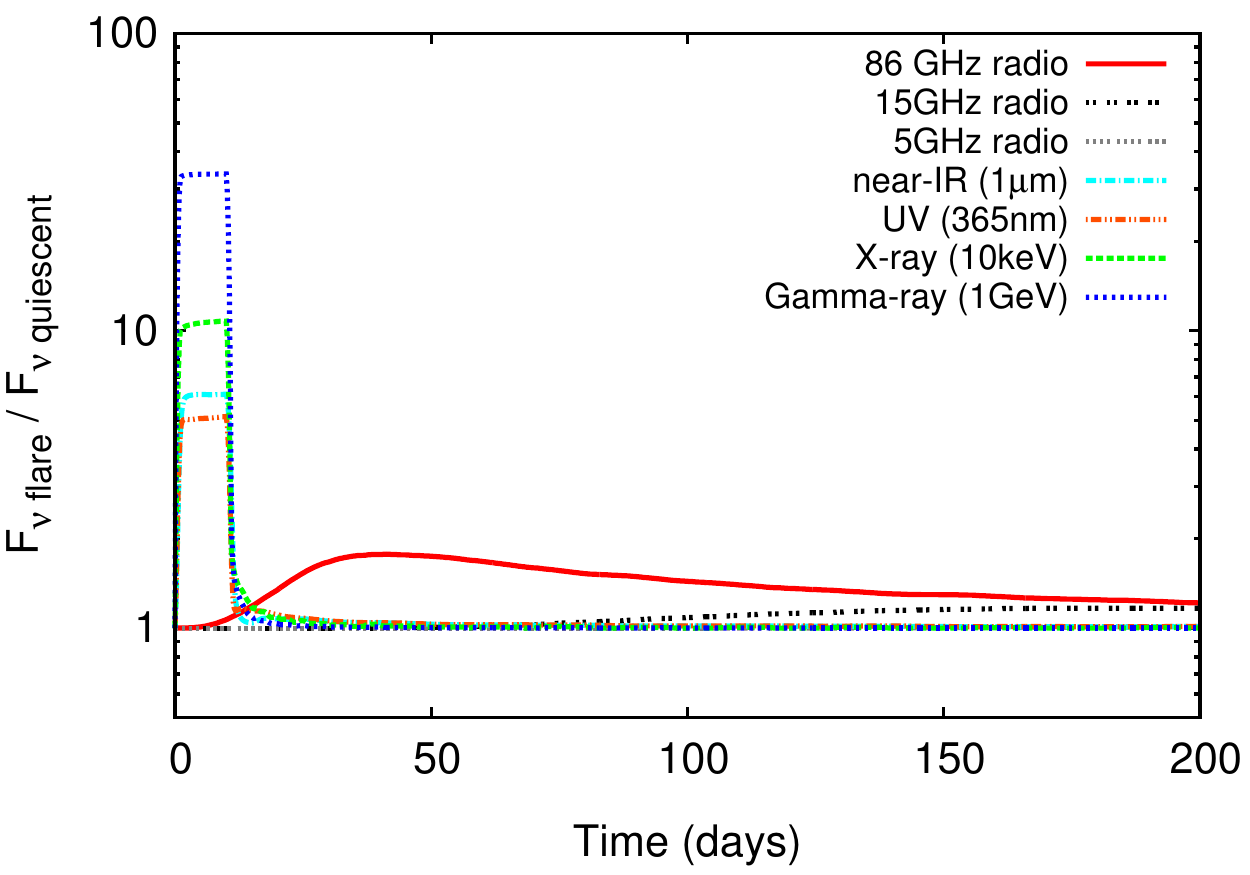} }
\vspace{-0.2cm}
		\subfloat[$x_{\m{flare}}=0.304\m{pc}$, \qquad$\Delta t_{\m{flare}}=$100 days]{ \includegraphics[width=6.8cm, clip=true, trim=0cm 0cm 0.cm 0cm]{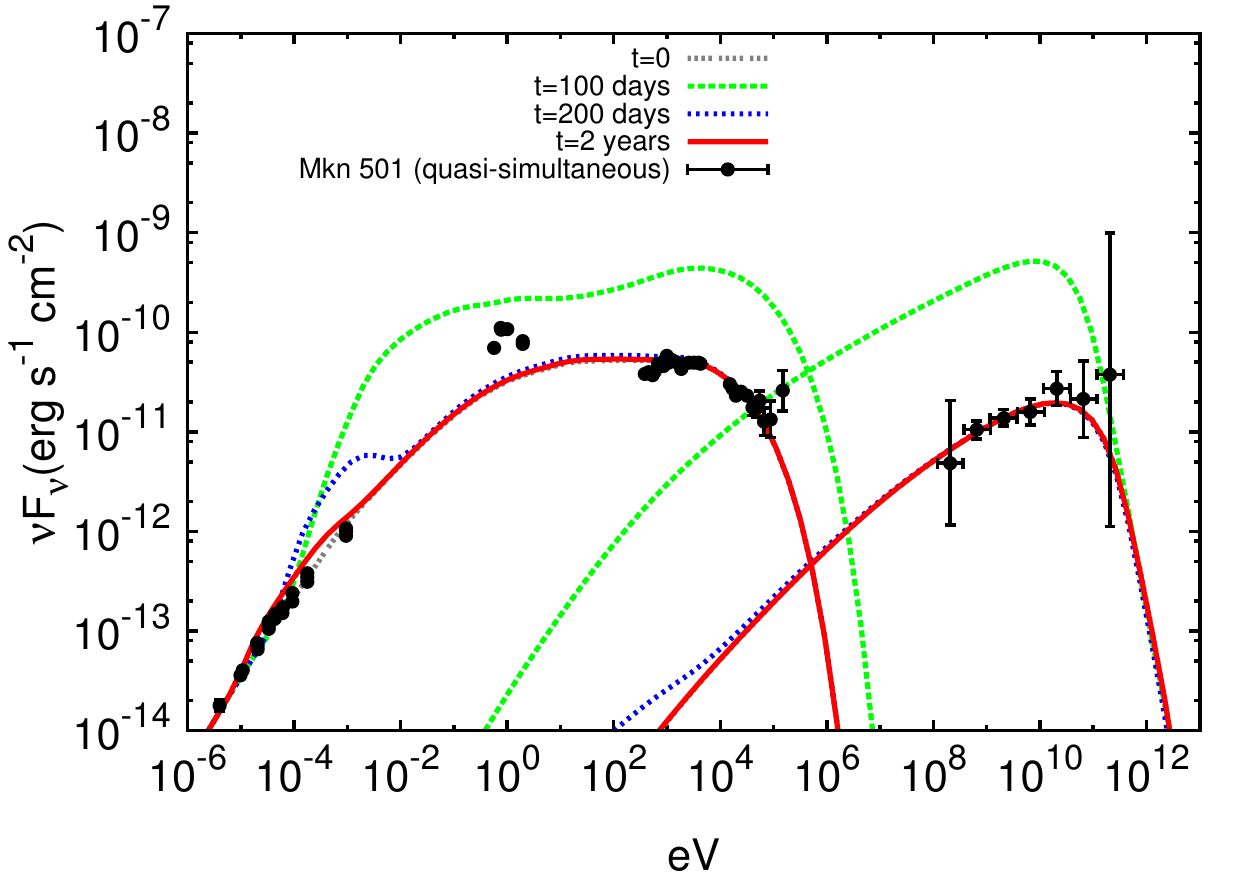} }
\qquad
		\subfloat[$x_{\m{flare}}=0.304\m{pc}$,\qquad $\Delta t_{\m{flare}}=$100 days]{ \includegraphics[width=6.8cm, clip=true, trim=0cm 0cm 0.cm 0cm]{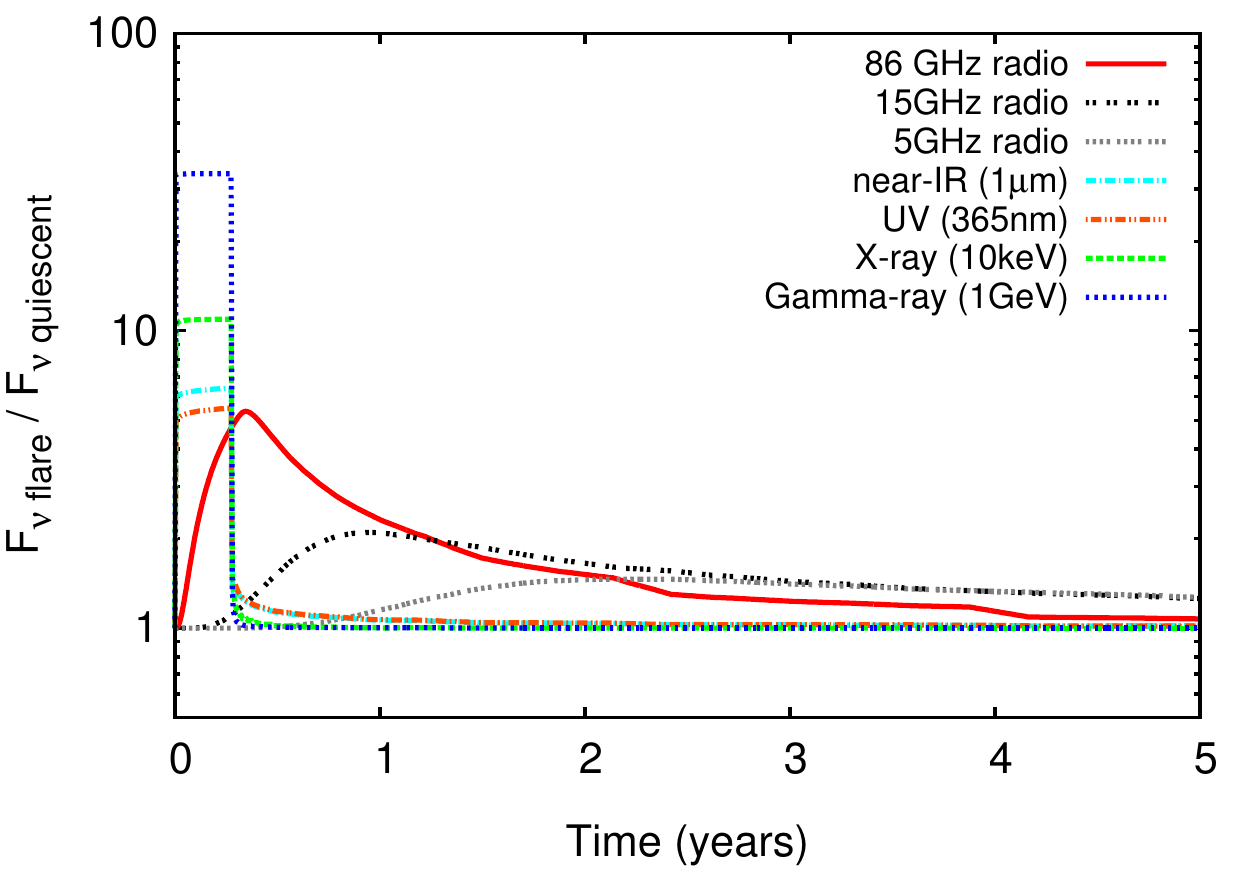} }
\vspace{-0.2cm}
		\subfloat[$x_{\m{flare}}=40.2\m{pc}$, \qquad$\Delta t_{\m{flare}}=$1000 days]{ \includegraphics[width=6.8cm, clip=true, trim=0cm 0cm 0.cm 0cm]{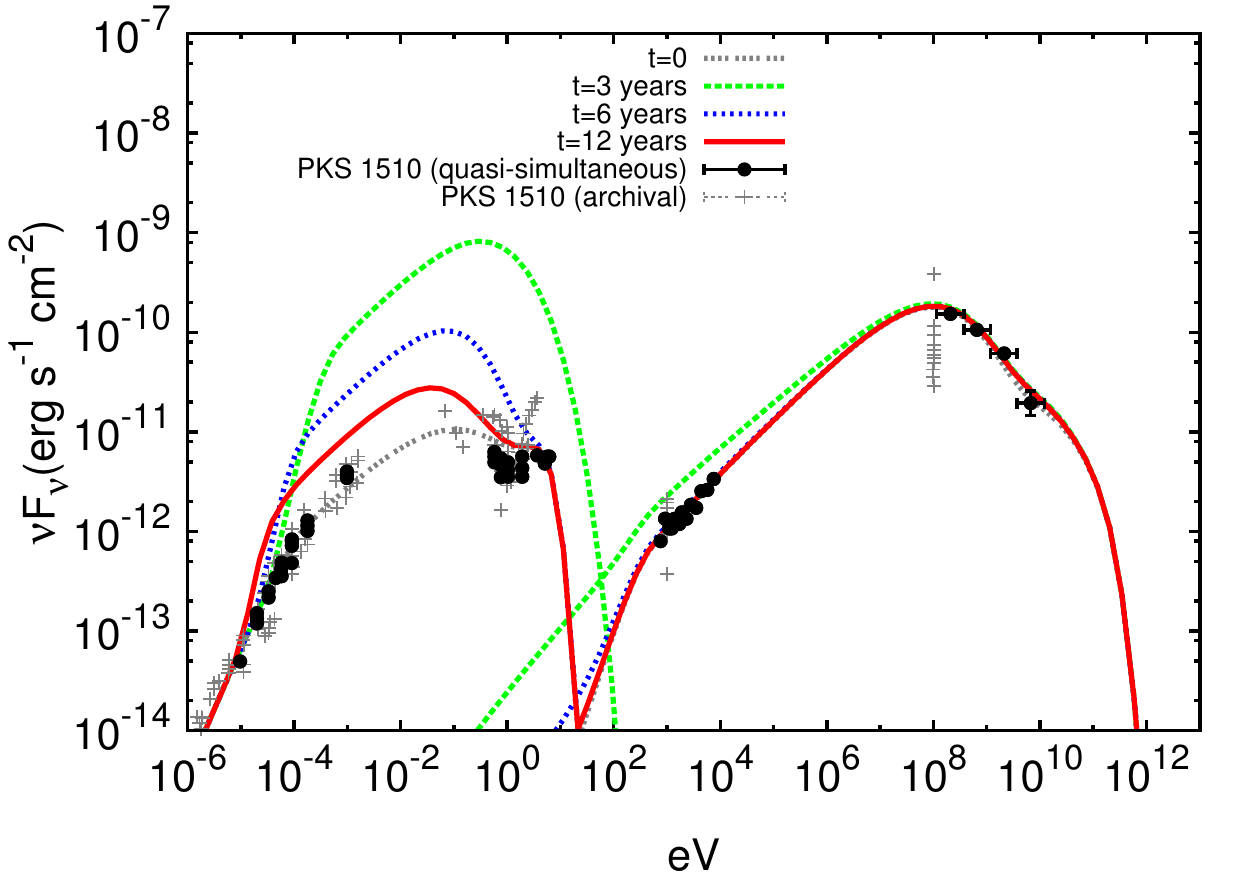} }
\qquad
		\subfloat[$x_{\m{flare}}=40.2\m{pc}$,\qquad $\Delta t_{\m{flare}}=$1000 days]{ \includegraphics[width=6.8cm, clip=true, trim=0cm 0cm 0.cm 0cm]{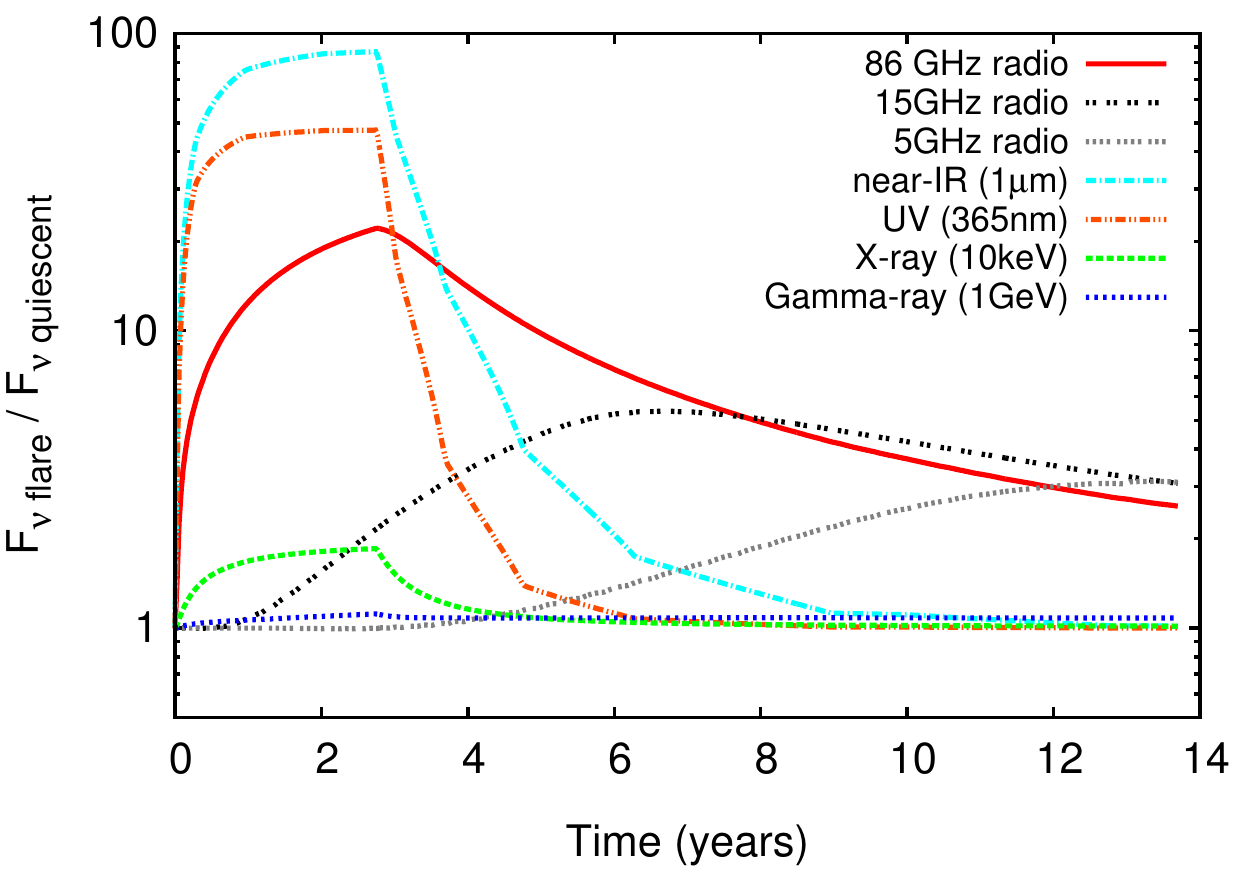} }
\vspace{-0.2cm}
		\subfloat[$x_{\m{flare}}=40.2\m{pc}$, \qquad$\Delta t_{\m{flare}}=10^{4}$ days]{ \includegraphics[width=6.8cm, clip=true, trim=0cm 0cm 0.cm 0cm]{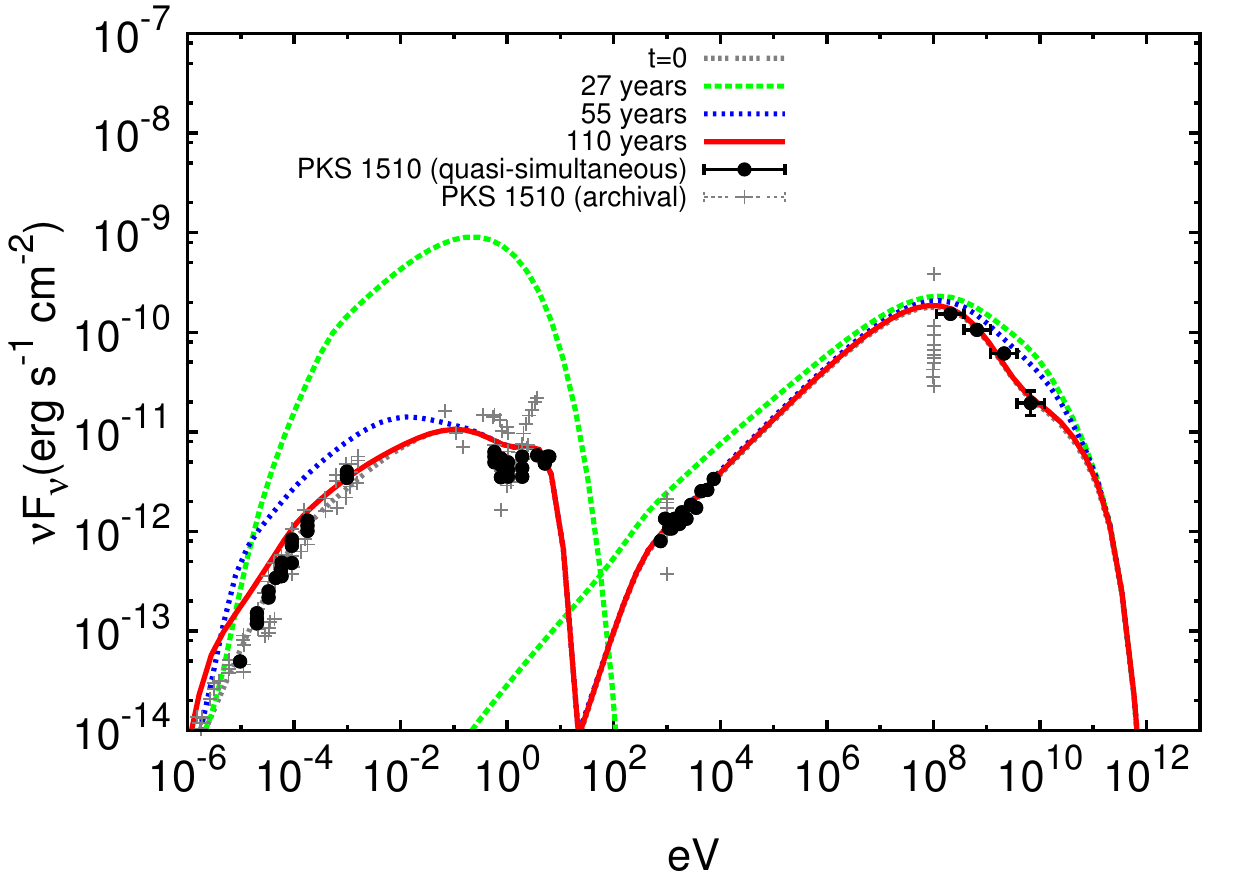} }
\qquad
		\subfloat[$x_{\m{flare}}=40.2\m{pc}$, \qquad$\Delta t_{\m{flare}}=10^{4}$ days]{ \includegraphics[width=6.8cm, clip=true, trim=0cm 0cm 0.cm 0cm]{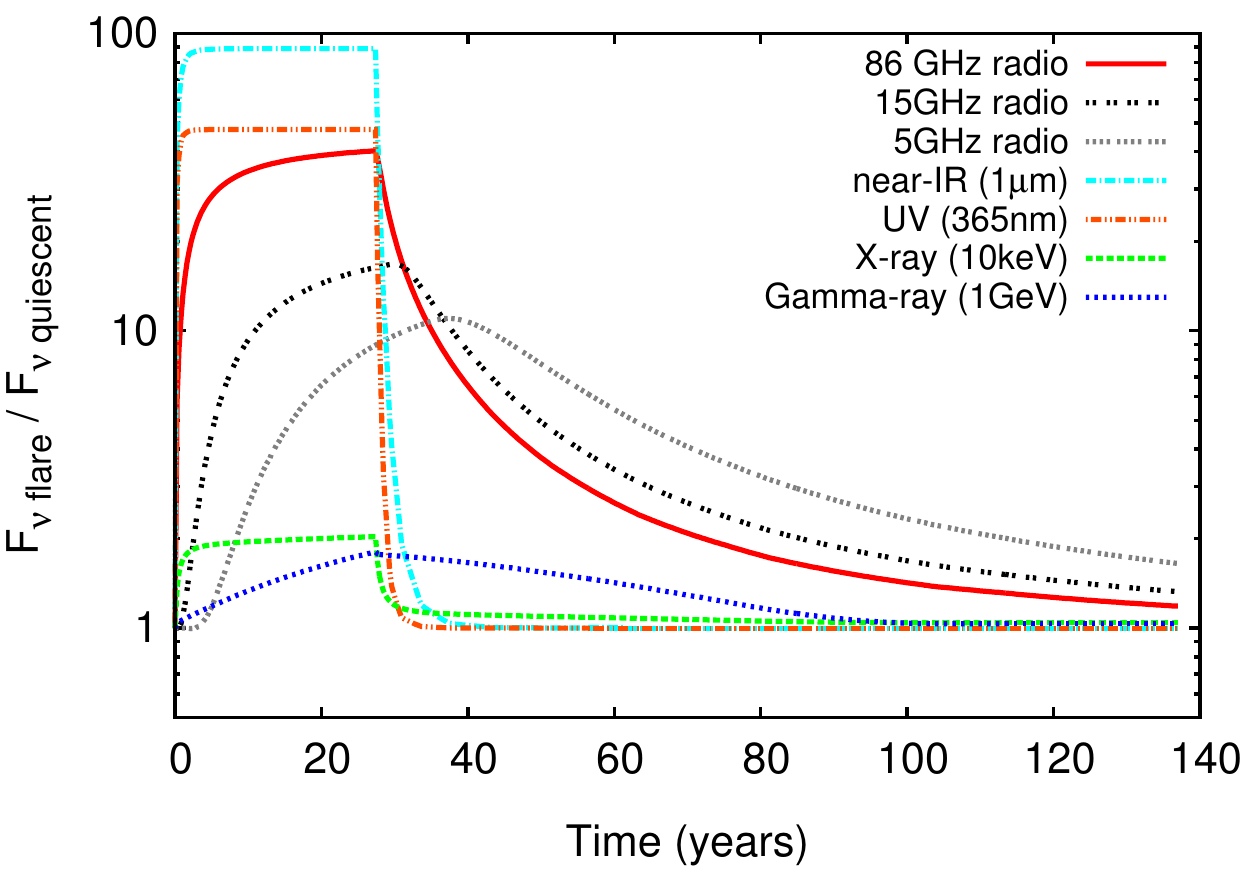} }	
		
	\caption{The effect of increasing the jet power by a factor of 10 for an observed duration given by $\Delta t_{\m{flare}}$, whilst keeping all other model parameters fixed. There is no difference between the flaring and quiescent models except for the jet power.  The high power plasma reaches the transition region where it first comes into equipartition at $t=0$. The time-dependent spectra and lightcurves show the lag between the prompt high energy emission and the optically thick radio emission. The lag increases at lower radio frequencies where the jet remains optically thick to larger distances. The total delay between the peak of the prompt flare at high energies and the peak of the flare at 5GHz increases substantially with jet power from $\sim$2 years in Mkn 501 to $\sim$20 years in the high power FSRQ PKS1510. The rise and decay timescales are substantially longer than at higher frequencies because the radiative lifetime of the radio emitting electrons is longer (see section \ref{radiosection} for discussion).  } 
\label{radio}
\end{figure*}

\begin{figure*}
	\centering
\begin{tabular}{cc}
\hspace{-0.4cm}
		\subfloat{ \includegraphics[width=7.3cm, clip=true, trim=0.cm 0cm 0.cm 0cm]{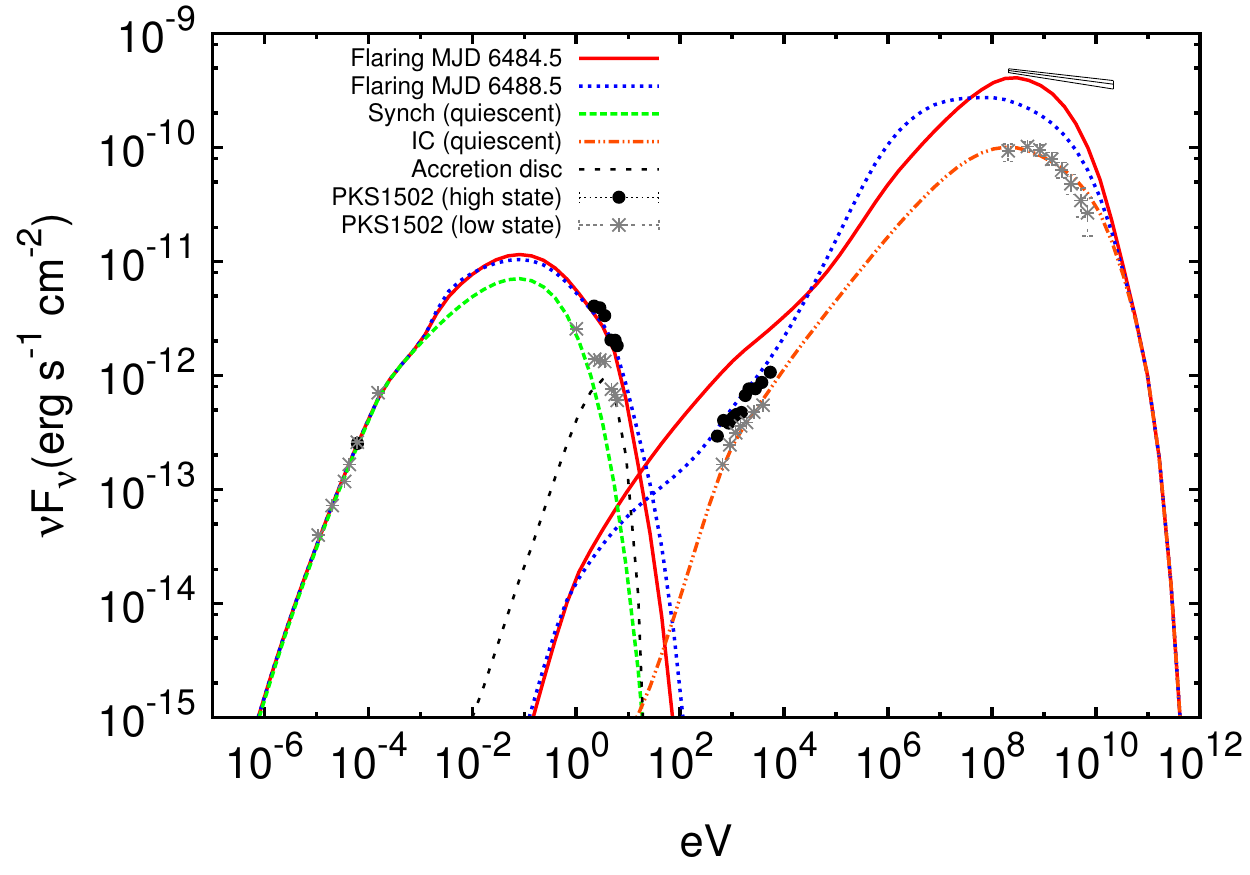} } &
		\multirow{-16}[0]{20cm}{\subfloat{ \includegraphics[width=15cm, clip=true, trim=0cm 0cm 0.cm 0cm]{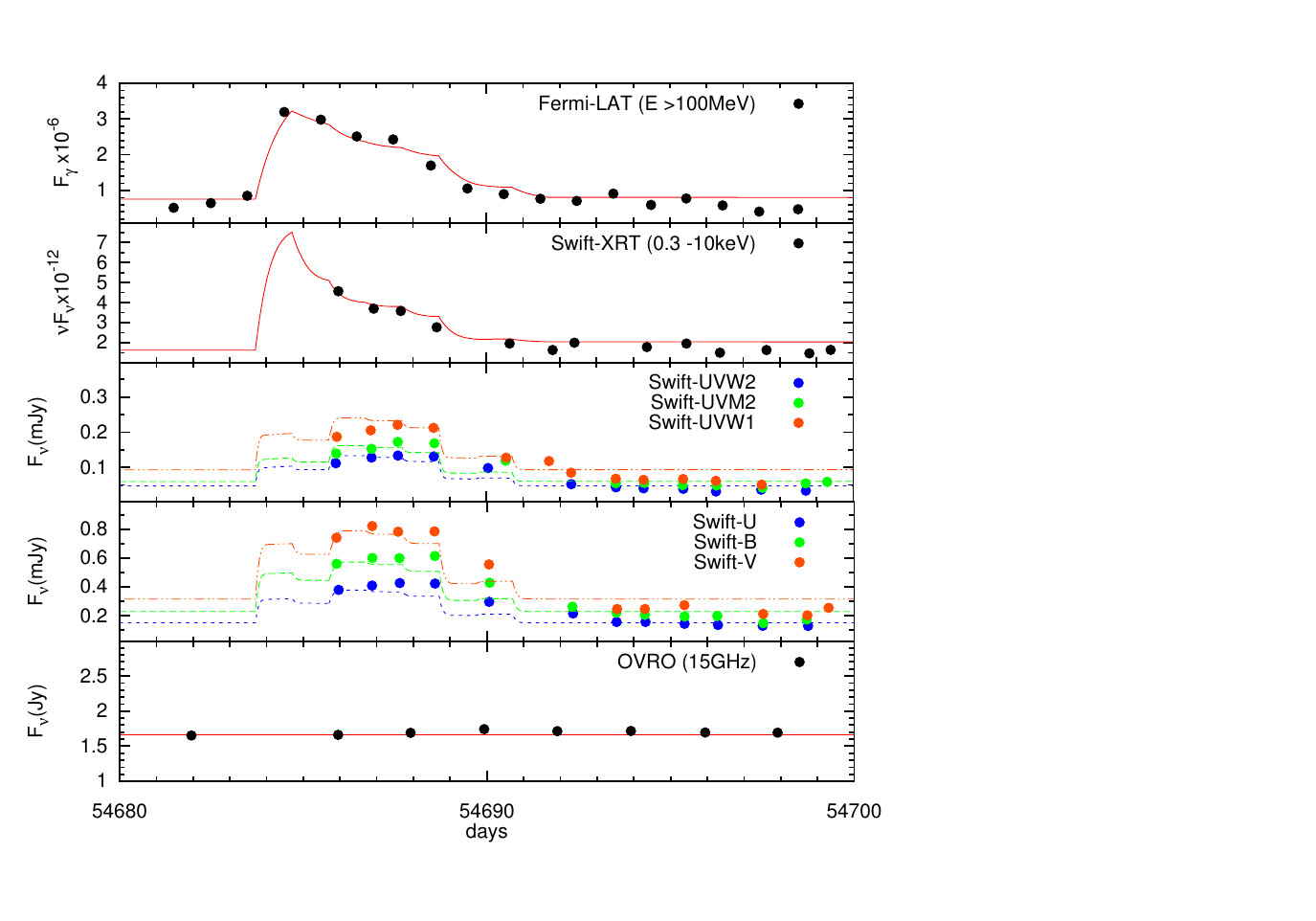} }}
\\
\hspace{-0.4cm}\subfloat{ \includegraphics[width=7.0cm, clip=true, trim=0cm 0cm 0.cm 0cm]{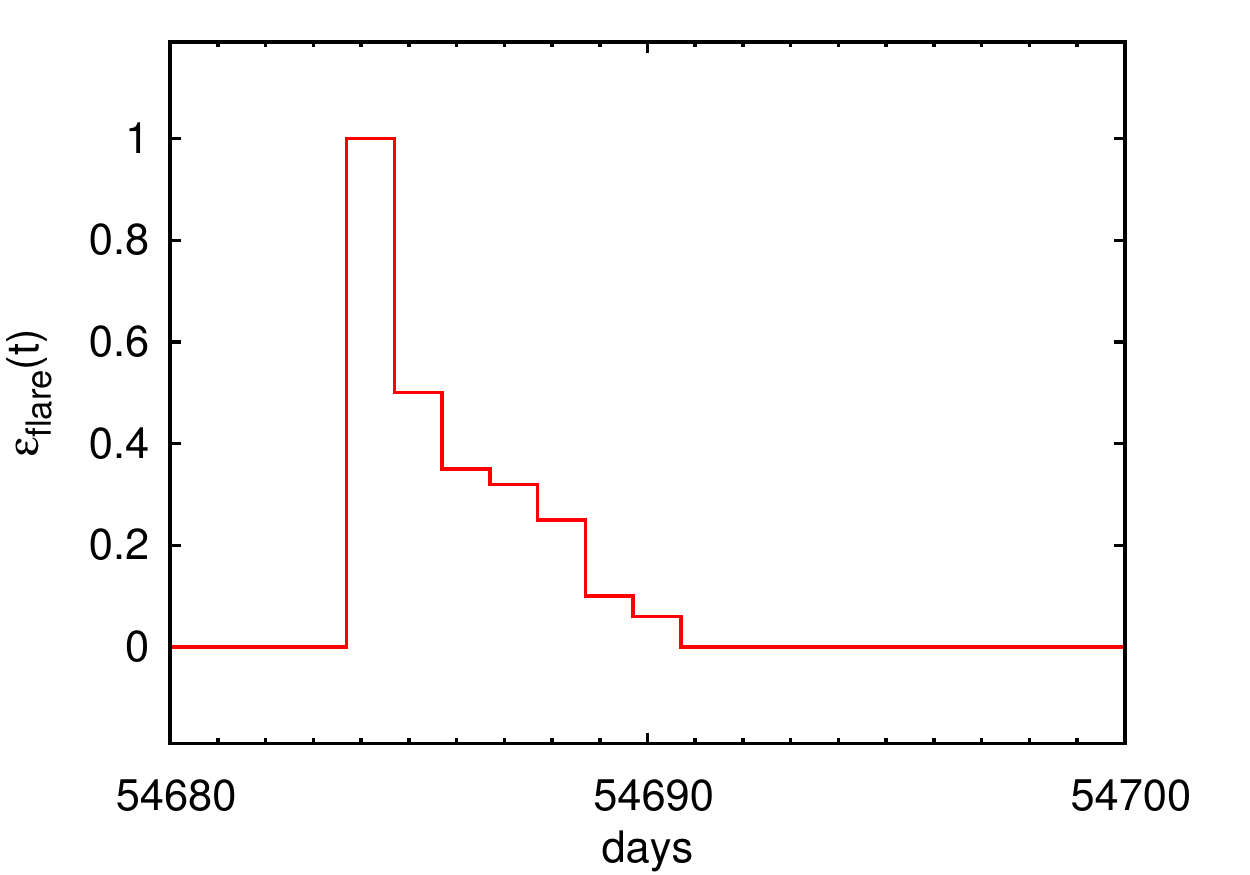} }
\end{tabular}
	\caption{Model fits to the observed spectrum and lightcurve of the $\sim$7 day flare in PKS1502 \citep{2010ApJ...710..810A}. The model fits well to the flare using only four free parameters: the distance of the flaring front $x_{\m{flare}}=0.312\m{pc}$, and the equipartition fraction of the plasma after passing through the flaring front as a function of time $\epsilon_{\m{flare}}$. The other properties of the flaring plasma are exactly the same as those of the quiescent plasma, which were determined by fitting to the quiescent spectrum (see Table \ref{Table2}). It is worth noting that the observed SED of the high state is not simultaneous. This is why our model SEDs do not fit simultaneously to both the X-ray and $\gamma$-ray high-state data.  } 
\label{fit}
\end{figure*}

\section{Radio flaring and the radio lag} \label{radiosection}

The calculations in the previous section showed that flares occurring close to the jet base, with flaring plasma that had the same intrinsic properties as the quiescent plasma, produced no observable luminosity changes at radio frequencies. This was because the jet plasma close to the jet base is optically thick to radio emission and the rise time of the flare (i.e. the radiative lifetime of radio emitting electrons) was very long compared to the short duration of the flares. In order to model and understand the properties of radio flares observed in jets, it is necessary to change the jet power as a function of time. Changing only the location at which the jet plasma first experiences strong particle acceleration and comes into equipartition does not have a large effect on the radio luminosity. This is because radio emitting electrons have a long radiative cooling lifetime and the low frequency radio emission will come primarily from large distances along the jet where the plasma becomes optically thin. The long radiative lifetime (weeks to decades) means that only comparatively long duration flaring events will be observed. Also, because both the quiescent and flaring plasma are close to equipartition at these large distances, the radio luminosity will depend primarily on the jet power of the radio emitting plasma. 

For these reasons we study the radio luminosity by changing the jet power between the quiescent ($W_{J\m{q}}$) and flaring models ($W_{J\m{f}}$), for an observed duration, $\Delta t_{\m{flare}}$, whilst leaving all other jet parameters unchanged. Figure \ref{radio} shows the results of changing the jet power for different durations. $t=0$, corresponds to the observed time at which the first flaring plasma passes the transition region of the jet, experiences particle acceleration and comes into equipartition, i.e. $\epsilon_{\m{flare}}=1$. The optically thin emission increases in luminosity with an intrinsic rise time given by its radiative lifetime smoothed by the geometrical time delay across the flaring front (\ref{geom}), i.e. the highest frequency synchrotron and inverse-Compton emission from the highest energy electrons, with the shortest lifetimes, rise fastest. The flaring luminosities then reach a plateau after their rise for the remaining duration of the flare, and then decay, also on the same timescale. We shall refer to the emission from these optically thin frequencies as \lq{}prompt emission\rq{}. There is a time-lag between the start of the rise of the prompt emission and the start of the rise of the optically thick radio emission. This lag is determined by the time taken for the flaring plasma to travel along the jet, to the distance at which the radio emission becomes optically thin. Since the jet plasma remains optically thick to larger distances for lower frequency radio emission, this lag increases with decreasing radio frequency. The radiative lifetime of the emitting electrons increase at lower radio frequencies meaning that the intrinsic rise and decay timescales also increase at lower radio frequencies. This combination of increasing lag and increasing rise and decay timescales as we go to lower radio frequencies means that the lowest observed radio frequencies have very substantial delays between the peak of the prompt high energy emission and the peak of the low frequency radio emission. At 5GHz this delay is approximately 2 years for Mkn501 and 20 years for PKS1510. Due to the long radiative lifetime of radio emitting electrons, radio lightcurves effectively show us how the long-term average jet power changes with time. Lower radio frequencies effectively show a moving average of the jet power smoothed over longer radiative timescales. 

It is important to note that at a given optically thick radio frequency the lag between prompt and radio emission depends on the jet power and geometry. This is because the lag is the travel time of the plasma from the optically thick transition region, responsible for the prompt emission, to the distance at which the jet becomes optically thin to radio emission.  This distance increases with jet power and for jets with a narrower opening angle. This means that we expect low power BL Lac type blazars to have shorter radio lags than high power FSRQs, with typical optically thick radio lags at 15GHz of $\sim$months  for BL Lacs and $\sim$years for FSRQs. Current radio monitoring campaigns have found a range of radio lags typically from $\sim$months -- years (\citealt{2014MNRAS.445..428M}, \citealt{2014MNRAS.441.1899F}, \citealt{2015MNRAS.452.1280R}), consistent with our results. Our results suggest that whilst a lag of $\sim$months may be typical for low power BL Lacs at 15GHz, high power FSRQ lags are likely to be substantially longer, especially at lower frequencies where lags can reach $\sim$decades. Such long lags would exceed the entire lifetime of most radio monitoring campaigns, making a statistical analysis very difficult. This indicates the need for continued long term radio monitoring in order to be able to determine blazar jet properties from observed radio lags. 

The rise and decay timescales at 15GHz are $\sim$ 6 months for Mkn501 and $\sim$ 5 years for PKS1510. The rise and decay timescales can become increasingly asymmetric at lower radio frequencies, as shown in Fig. \ref{radio}. This is because at low radio frequencies the radiative lifetime of the emitting electrons becomes so long that the flaring plasma can travel substantial distances along the jet. These distances are large enough that the jet radius can substantially increase and magnetic field decrease, causing the synchrotron radiative lifetime of the electrons to increase with time. This means that the decay timescale of the flare can exceed the rise timescale because the radiative lifetime is increasing along the jet. We predict that at lower radio frequencies the asymmetry between rise and decay timescales should become more pronounced as the emitting electrons are able to travel even larger distances along the jet. 

These new results are particularly significant because this is the first time that optically thick radio flares and radio lags have been properly modelled and related to the physical properties of the jet.

\section{Fitting to observations} \label{section6}

In the previous sections we calculated how the characteristic properties of blazar flares are determined by the jet properties and location of the flaring event. In order to test whether our model is physically realistic and appropriate, it is important that the model is able to fit to observations of a real flaring event. For this purpose we choose to fit the 2008 flare of PKS$1502+106$ \citep{2010ApJ...710..810A}, which was observed at multiple wavelengths simultaneously and has a similar lightcurve profile at multiple wavelengths (making it unlikely that the simultaneous flares at multiple wavelengths were unrelated stochastic events). In order to limit the number of free parameters of our model fit, the only properties which we allow to change between the flaring plasma and quiescent plasma are the location of the flaring front, $x_{\m{flare}}$, where the flaring plasma experiences strong particle acceleration, the jet power of the flaring plasma $W_{\m{J f}}$, the time-dependent equipartition fraction of plasma passing through the flaring front, $\epsilon_{\m{flare}}(t)$, and the maximum electron energy of the injected electron spectrum in the flare $E_{\m{max\ f}}$. In previous sections the time-dependent equipartition fraction of the flaring plasma passing through the flaring front, $\epsilon_{\m{flare}}(t)$, was chosen to be a rectangular or \lq{}top hat\rq{} function, for illustrative purposes. For a realistic flare, however, we expect a continuous increase and decrease in the particle acceleration power at the flaring front and so here we allow the equipartition fraction of the plasma passing through the flaring front to change as a function of time. Since the cadence of the observations is approximately 1 day, the value of $\epsilon_{\m{flare}}$ is also chosen to change with this cadence.  

The model fit to the observed flaring event and SED is shown in Figure \ref{fit} and the flaring model parameters in Table \ref{Table1}. Using only four parameters to describe the flaring event (three constants $x_{\m{flare}}$, $W_{\m{J f}}$ and $E_{\m{max\ f}}$ and one time-dependent variable $\epsilon_{\m{flare}}$), the model fits accurately both to the observed time-dependent spectrum and the nine multiwavelength lightcurves across most wavelengths. In particular, the radio, UV, X-ray and $\gamma$-ray data are fitted well, however, we find it difficult to fit the precise shape of the decaying optical lightcurves simultaneously with the high energy emission, using only four free parameters (the decaying optical data is overproduced by $\sim50\%$ in our fit). The good fit across most wavelengths is strong evidence that the model captures the key physical properties of the observed flare, because the majority of properties of the flaring plasma (such as the jet shape and bulk Lorentz factor as a function of distance along the jet) are left identical to those determined by our fit to the quiescent jet spectrum (see Table \ref{Table2}). This consistency between the flaring and quiescent plasma parameters implies that the basic physical assumptions of our flaring model accurately reflect the processes involved in real jets. It is worth noting that the two parameters which are changed from the quiescent fit: the flaring jet power $W_{\m{J f}}$ and maximum electron energy $E_{\m{max\ f}}$ do not differ dramatically from the quiescent parameters, changing by $45\%$ and $46\%$ respectively. Since we know blazar jets to be highly variable, we would expect flaring jet parameters to differ slightly from their quiescent values and so these changes in parameters are not surprising.

The location of the flaring front in this fit is at a distance of $x_{\m{flare}}=0.31\m{pc}$ along the jet. This is in the parabolic section of the jet, inside the BLR (which extends to $\sim$0.5pc \citealt{2005ApJ...629...61K}, \citealt{2013MNRAS.429.1189P}) and much closer to the jet base than the location of the transition region in our quiescent fit, $x_{\m{T}}=29$pc, the location at which the quiescent jet plasma first experiences strong particle acceleration and comes into equipartition. The model fit suggests that this flare was caused by a transient particle acceleration event occurring in the parabolic section of the jet, as opposed to being caused by a dramatic time-dependent change in the properties of the jet plasma, such as the jet power. In quiescence, the magnetised parabolic section of the jet is highly magnetised and therefore not very luminous \citep{2017MNRAS.465..337P}. It will be important to understand whether the majority of blazar flares are caused by transient particle acceleration events occurring in the magnetised parabolic jet base (as in this case), dramatic time-dependent changes in the properties of the jet plasma, or a combination of both. This question will be addressed in future work by fitting our model to multiple observations of flaring events. 

The multiwavelength lightcurve of PKS1502 shows complex behaviour. In particular, whilst the $\gamma$-ray and X-ray flare seem to coincide and have similar flaring profiles, the observed peak of the optical/UV flare is delayed by several days with respect to the $\gamma$-ray/X-ray peak (although this does not preclude a prompt optical/UV peak occurring close to the start of the flare, before the Swift observations began). Our model is able to fit to this delayed rise at optical/UV wavelengths, which we find to be caused by the changing equipartition fraction of the freshly accelerated flaring plasma. At the start of the flare, the flaring front initially accelerates plasma to equipartition, which is optimal for producing SSC emission. As the flare progresses, the power injected into accelerating particles at the flaring front decreases and so the plasma accelerated by the flaring front becomes more magnetised (as can be seen from $\epsilon_{\m{flare}}$ in Fig. \ref{fit}). This results in less SSC emission and thereby a slight increase in synchrotron emission, causing the delayed rise at optical/UV frequencies. 

There is no radio flare associated with this event for two reasons: the total energy density of the flaring plasma is close to that of the quiescent plasma and so at large distances it will have a similar radio luminosity; the duration of the flare is short compared to the radiative lifetime of radio emitting electrons and so is much shorter that than the characteristic rise and decay timescales at radio frequencies. This is consistent with the results of section \ref{radiosection} which conclude that short duration flares ($<$week) at 15GHz are unlikely to be luminous enough to be observed at radio frequencies due to the long rise and decay timescales.

\section{Conclusions}

In this paper we have developed a time-dependent inhomogeneous jet fluid emission model in order to study the properties of blazar flares. The model allows us to study transient particle acceleration events occurring anywhere along the extended jet structure, to understand how the location and duration of a flare affects the observed spectrum and lightcurve. Our model calculates the fluid evolution of flaring plasma as it propagates along the jet, conserving relativistic energy-momentum, unlike previous static spherical blob models. The model incorporates a realistic large scale parabolic to conical jet structure.  This allows us to model the expected radio time lag associated with a high energy flaring event for the first time. The flaring events occur in an otherwise quiescent jet plasma, which is responsible for producing the observed quiescent emission.

The model was used to calculate the time-dependent spectra and lightcurves produced by transient particle acceleration events occurring at a variety of distances along typical BL Lac and FSRQ type blazar jets. The location of the flare leaves a distinctive signature on the observed spectrum and lightcurves allowing the location of the flare to be determined from observations. We tested the physical applicability of our model by fitting to the observed multiwavelength lightcurves and SED of the 2008 flare of FSRQ PKS1502. The flare is accurately fitted by the model using a transient particle acceleration event occurring in the parabolic base of the jet, at a distance of 0.31pc, inside the BLR. Significantly, we find that the majority of the properties of the flaring plasma in this fit are identical to the quiescent plasma. This implies that the main physical assumptions of our flaring model are accurate.

From this model we learn several key points:
\begin{itemize}
\item The intrinsic rise and decay timescales of the luminosity of a flare are both given approximately by the radiative lifetime of the electrons emitting at a given frequency smoothed by the geometrical path length differences between the observer and different points along the flaring front (see fig \ref{schematic3} for a detailed explanation). 
\item This leads to a general prediction that flares come in two general categories: {\bf symmetric flares} - approximately symmetric flares, for flaring events whose duration is shorter than both the radiative lifetime of the emitting particles and the geometric delay timescale; {\bf extended flares} - flares which possess an extended structure tracking the variations in the power of the flaring particle acceleration event. This is the case for flares whose duration exceeds both the radiative lifetime of the emitting particles and the geometric delay timescale. 
\item Orphan flares are a common phenomenon which occur because a flare only becomes visible at a particular frequency when its luminosity exceeds the quiescent emission at that frequency. Since the characteristic rise time in luminosity is shortest for the highest energy synchrotron and inverse-Compton emission, we expect X-ray and $\gamma$-ray orphan flares to be the most commonly observed. The Compton-dominance of the flaring region determines whether an orphan flare is an X-ray or $\gamma$-ray flare.  
\item Flaring at optically thick radio frequencies is usually delayed with respect to prompt optically thin flaring emission. The delay, or radio lag, is determined by the observed travel time of the flaring plasma from the initial flare location out to large distances where the radio emission becomes optically thin and brightest. The radio lag increases with jet power because in higher power jets the radio emitting regions occur at larger distances than in lower power jets. For low power BL Lac blazars, such as Markarian 501, the radio lag at 5GHz is roughly $\sim 6$ months, whilst in the high power FSRQ PKS1510 this lag increases to $\sim 4$ years. This lag increases at lower radio frequencies because at lower frequencies the jet remains optically thick to larger distances resulting in longer travel times. 
\item Due to the long radiative lifetime of radio emitting electrons radio flares effectively track much longer timescale variations in the jet power than the prompt high energy emission. This means that short $\gamma$-ray flares do not contain enough energy to be observable above the quiescent radio emission. Only changes in the jet power which are comparable to, or longer than, the radio emitting lifetime of electrons are observable. The total delay between the peak of the optically thin prompt flaring emission and the peak luminosity of the flare at 15GHz is $\sim$ 1 year for the low power blazar Markarian 501 and $\sim$ 7 years for the high power FSRQ PKS1510. 
\end{itemize} 
These theoretical predictions can be tested against observations to check the accuracy and assumptions of our model. 

This work demonstrates the potential for realistic inhomogeneous fluid jet models to be used to better understand the physical processes responsible for multiwavelength blazar flares. 

\section*{Acknowledgements}

WJP acknowledges funding from a Junior Research Fellowship from University College, University of Oxford. WJP would like to thank Garret Cotter and Paul Morris for useful discussions surrounding blazar flares, and the anonymous referee for many helpful suggestions and comments. 

\bibliographystyle{mn2e}
\bibliography{References}

\begin{thebibliography}{}

\bibitem[\protect\citeauthoryear{{Abdo}, {Ackermann}, {Ajello}, {Atwood},
  {Axelsson}, {Baldini}, {Ballet}, {Barbiellini}, {Bastieri}, {Baughman} \& et
  al.}{{Abdo} et~al.}{2010}]{2010ApJ...710..810A}
{Abdo} A.~A.,  {Ackermann} M.,  {Ajello} M.,  {Atwood} W.~B.,  {Axelsson} M.,
  {Baldini} L.,  {Ballet} J.,  {Barbiellini} G.,  {Bastieri} D.,  {Baughman}
  B.~M.,    et al. 2010, \apj, 710, 810

\bibitem[\protect\citeauthoryear{{Acciari}, {Aliu}, {Arlen}, {Aune},
  {Beilicke}, {Benbow}, {Boltuch}, {Bradbury}, {Buckley}, {Bugaev} \& et
  al.}{{Acciari} et~al.}{2011}]{2011ApJ...738...25A}
{Acciari} V.~A.,  {Aliu} E.,  {Arlen} T.,  {Aune} T.,  {Beilicke} M.,  {Benbow}
  W.,  {Boltuch} D.,  {Bradbury} S.~M.,  {Buckley} J.~H.,  {Bugaev} V.,    et
  al. 2011, \apj, 738, 25

\bibitem[\protect\citeauthoryear{{Acciari}, {Arlen}, {Aune}, {Beilicke},
  {Benbow}, {B{\"o}ttcher}, {Boltuch}, {Bradbury}, {Buckley}, {Bugaev} \& et
  al.}{{Acciari} et~al.}{2011}]{2011ApJ...729....2A}
{Acciari} V.~A.,  {Arlen} T.,  {Aune} T.,  {Beilicke} M.,  {Benbow} W.,
  {B{\"o}ttcher} M.,  {Boltuch} D.,  {Bradbury} S.~M.,  {Buckley} J.~H.,
  {Bugaev} V.,    et al. 2011, \apj, 729, 2

\bibitem[\protect\citeauthoryear{{Aharonian}, {Akhperjanian}, {Bazer-Bachi},
  {Behera}, {Beilicke}, {Benbow}, {Berge}, {Bernl{\"o}hr}, {Boisson}, {Bolz} \&
  et al.}{{Aharonian} et~al.}{2007}]{2007ApJ...664L..71A}
{Aharonian} F.,  {Akhperjanian} A.~G.,  {Bazer-Bachi} A.~R.,  {Behera} B.,
  {Beilicke} M.,  {Benbow} W.,  {Berge} D.,  {Bernl{\"o}hr} K.,  {Boisson} C.,
  {Bolz} O.,    et al. 2007, \apjl, 664, L71

\bibitem[\protect\citeauthoryear{{Albert}, {Aliu}, {Anderhub}, {Antoranz},
  {Armada}, {Baixeras}, {Barrio}, {Bartko}, {Bastieri}, {Becker}, {Bednarek},
  {Berger} \& et al.}{{Albert} et~al.}{2007}]{2007ApJ...669..862A}
{Albert} J.,  {Aliu} E.,  {Anderhub} H.,  {Antoranz} P.,  {Armada} A.,
  {Baixeras} C.,  {Barrio} J.~A.,  {Bartko} H.,  {Bastieri} D.,  {Becker}
  J.~K.,  {Bednarek} W.,  {Berger} K.,    et al. 2007, \apj, 669, 862

\bibitem[\protect\citeauthoryear{{Aleksi{\'c}}, {Alvarez}, {Antonelli},
  {Antoranz}, {Asensio}, {Backes}, {Barrio}, {Bastieri}, {Becerra
  Gonz{\'a}lez}, {Bednarek} \& et al.}{{Aleksi{\'c}}
  et~al.}{2012}]{2012A&A...542A.100A}
{Aleksi{\'c}} J.,  {Alvarez} E.~A.,  {Antonelli} L.~A.,  {Antoranz} P.,
  {Asensio} M.,  {Backes} M.,  {Barrio} J.~A.,  {Bastieri} D.,  {Becerra
  Gonz{\'a}lez} J.,  {Bednarek} W.,    et al. 2012, \aap, 542, A100

\bibitem[\protect\citeauthoryear{{Asada} \& {Nakamura}}{{Asada} \&
  {Nakamura}}{2012}]{2012ApJ...745L..28A}
{Asada} K.,  {Nakamura} M.,  2012, \apjl, 745, L28

\bibitem[\protect\citeauthoryear{{Balokovi{\'c}}, {Paneque}, {Madejski},
  {Furniss}, {Chiang}, {Ajello}, {Alexander}, {Barret}, {Blandford}, {Boggs} \&
  et al.}{{Balokovi{\'c}} et~al.}{2016}]{2016ApJ...819..156B}
{Balokovi{\'c}} M.,  {Paneque} D.,  {Madejski} G.,  {Furniss} A.,  {Chiang} J.,
   {Ajello} M.,  {Alexander} D.~M.,  {Barret} D.,  {Blandford} R.~D.,  {Boggs}
  S.~E.,    et al. 2016, \apj, 819, 156

\bibitem[\protect\citeauthoryear{{Bartoli}, {Bernardini}, {Bi}, {Cao},
  {Catalanotti}, {Chen}, {Chen}, {Cui}, {Dai}, {D'Amone} \& et al.}{{Bartoli}
  et~al.}{2016}]{2016ApJS..222....6B}
{Bartoli} B.,  {Bernardini} P.,  {Bi} X.~J.,  {Cao} Z.,  {Catalanotti} S.,
  {Chen} S.~Z.,  {Chen} T.~L.,  {Cui} S.~W.,  {Dai} B.~Z.,  {D'Amone} A.,    et
  al. 2016, \apjs, 222, 6

\bibitem[\protect\citeauthoryear{{Bell}, {Schure} \& {Reville}}{{Bell}
  et~al.}{2011}]{2011MNRAS.tmp.1506B}
{Bell} A.~R.,  {Schure} K.~M.,    {Reville} B.,  2011, \mnras, p.~1506

\bibitem[\protect\citeauthoryear{{Blandford} \& {Znajek}}{{Blandford} \&
  {Znajek}}{1977}]{1977MNRAS.179..433B}
{Blandford} R.~D.,  {Znajek} R.~L.,  1977, \mnras, 179, 433

\bibitem[\protect\citeauthoryear{{B{\l}a{\.z}ejowski}, {Blaylock}, {Bond},
  {Bradbury}, {Buckley}, {Carter-Lewis}, {Celik}, {Cogan}, {Cui}, {Daniel} \&
  et al.}{{B{\l}a{\.z}ejowski} et~al.}{2005}]{2005ApJ...630..130B}
{B{\l}a{\.z}ejowski} M.,  {Blaylock} G.,  {Bond} I.~H.,  {Bradbury} S.~M.,
  {Buckley} J.~H.,  {Carter-Lewis} D.~A.,  {Celik} O.,  {Cogan} P.,  {Cui} W.,
  {Daniel} M.,    et al. 2005, \apj, 630, 130

\bibitem[\protect\citeauthoryear{{Bonnoli}, {Ghisellini}, {Foschini},
  {Tavecchio} \& {Ghirlanda}}{{Bonnoli} et~al.}{2011}]{2011MNRAS.410..368B}
{Bonnoli} G.,  {Ghisellini} G.,  {Foschini} L.,  {Tavecchio} F.,    {Ghirlanda}
  G.,  2011, \mnras, 410, 368

\bibitem[\protect\citeauthoryear{{B{\"o}ttcher} \& {Chiang}}{{B{\"o}ttcher} \&
  {Chiang}}{2002}]{2002ApJ...581..127B}
{B{\"o}ttcher} M.,  {Chiang} J.,  2002, \apj, 581, 127

\bibitem[\protect\citeauthoryear{{Broderick} \& {Tchekhovskoy}}{{Broderick} \&
  {Tchekhovskoy}}{2015}]{2015ApJ...809...97B}
{Broderick} A.~E.,  {Tchekhovskoy} A.,  2015, \apj, 809, 97

\bibitem[\protect\citeauthoryear{{Elitzur}}{{Elitzur}}{2006}]{2006NewAR..50..7%
28E}
{Elitzur} M.,  2006, \nar, 50, 728

\bibitem[\protect\citeauthoryear{{Finke}, {Dermer} \& {B{\"o}ttcher}}{{Finke}
  et~al.}{2008}]{2008ApJ...686..181F}
{Finke} J.~D.,  {Dermer} C.~D.,    {B{\"o}ttcher} M.,  2008, \apj, 686, 181

\bibitem[\protect\citeauthoryear{{Fossati}, {Maraschi}, {Celotti}, {Comastri}
  \& {Ghisellini}}{{Fossati} et~al.}{1998}]{1998MNRAS.299..433F}
{Fossati} G.,  {Maraschi} L.,  {Celotti} A.,  {Comastri} A.,    {Ghisellini}
  G.,  1998, \mnras, 299, 433

\bibitem[\protect\citeauthoryear{{Fuhrmann}, {Larsson}, {Chiang}, {Angelakis},
  {Zensus}, {Nestoras}, {Krichbaum}, {Ungerechts}, {Sievers}, {Pavlidou},
  {Readhead}, {Max-Moerbeck} \& {Pearson}}{{Fuhrmann}
  et~al.}{2014}]{2014MNRAS.441.1899F}
{Fuhrmann} L.,  {Larsson} S.,  {Chiang} J.,  {Angelakis} E.,  {Zensus} J.~A.,
  {Nestoras} I.,  {Krichbaum} T.~P.,  {Ungerechts} H.,  {Sievers} A.,
  {Pavlidou} V.,  {Readhead} A.~C.~S.,  {Max-Moerbeck} W.,    {Pearson} T.~J.,
  2014, \mnras, 441, 1899

\bibitem[\protect\citeauthoryear{{Goldreich} \& {Julian}}{{Goldreich} \&
  {Julian}}{1969}]{1969ApJ...157..869G}
{Goldreich} P.,  {Julian} W.~H.,  1969, \apj, 157, 869

\bibitem[\protect\citeauthoryear{{Kaspi}, {Maoz}, {Netzer}, {Peterson},
  {Vestergaard} \& {Jannuzi}}{{Kaspi} et~al.}{2005}]{2005ApJ...629...61K}
{Kaspi} S.,  {Maoz} D.,  {Netzer} H.,  {Peterson} B.~M.,  {Vestergaard} M.,
  {Jannuzi} B.~T.,  2005, \apj, 629, 61

\bibitem[\protect\citeauthoryear{{Kirk}, {Rieger} \& {Mastichiadis}}{{Kirk}
  et~al.}{1998}]{1998A&A...333..452K}
{Kirk} J.~G.,  {Rieger} F.~M.,    {Mastichiadis} A.,  1998, \aap, 333, 452

\bibitem[\protect\citeauthoryear{{Komissarov}, {Barkov}, {Vlahakis} \&
  {K{\"o}nigl}}{{Komissarov} et~al.}{2007}]{2007MNRAS.380...51K}
{Komissarov} S.~S.,  {Barkov} M.~V.,  {Vlahakis} N.,    {K{\"o}nigl} A.,  2007,
  \mnras, 380, 51

\bibitem[\protect\citeauthoryear{{Krawczynski}, {Hughes}, {Horan}, {Aharonian},
  {Aller}, {Aller}, {Boltwood} \& et al.}{{Krawczynski}
  et~al.}{2004}]{2004ApJ...601..151K}
{Krawczynski} H.,  {Hughes} S.~B.,  {Horan} D.,  {Aharonian} F.,  {Aller}
  M.~F.,  {Aller} H.,  {Boltwood} P.,    et al. 2004, \apj, 601, 151

\bibitem[\protect\citeauthoryear{{Li} \& {Kusunose}}{{Li} \&
  {Kusunose}}{2000}]{2000ApJ...536..729L}
{Li} H.,  {Kusunose} M.,  2000, \apj, 536, 729

\bibitem[\protect\citeauthoryear{{Marscher}}{{Marscher}}{2014}]{2014ApJ...780.%
..87M}
{Marscher} A.~P.,  2014, \apj, 780, 87

\bibitem[\protect\citeauthoryear{{Max-Moerbeck}, {Hovatta}, {Richards}, {King},
  {Pearson}, {Readhead}, {Reeves}, {Shepherd}, {Stevenson}, {Angelakis},
  {Fuhrmann}, {Grainge}, {Pavlidou}, {Romani} \& {Zensus}}{{Max-Moerbeck}
  et~al.}{2014}]{2014MNRAS.445..428M}
{Max-Moerbeck} W.,  {Hovatta} T.,  {Richards} J.~L.,  {King} O.~G.,  {Pearson}
  T.~J.,  {Readhead} A.~C.~S.,  {Reeves} R.,  {Shepherd} M.~C.,  {Stevenson}
  M.~A.,  {Angelakis} E.,  {Fuhrmann} L.,  {Grainge} K.~J.~B.,  {Pavlidou} V.,
  {Romani} R.~W.,    {Zensus} J.~A.,  2014, \mnras, 445, 428

\bibitem[\protect\citeauthoryear{{McKinney}}{{McKinney}}{2006}]{2006MNRAS.368.%
1561M}
{McKinney} J.~C.,  2006, \mnras, 368, 1561

\bibitem[\protect\citeauthoryear{{Nalewajko}, {Sikora}, {Madejski}, {Exter},
  {Szostek}, {Szczerba}, {Kidger} \& {Lorente}}{{Nalewajko}
  et~al.}{2012}]{2012ApJ...760...69N}
{Nalewajko} K.,  {Sikora} M.,  {Madejski} G.~M.,  {Exter} K.,  {Szostek} A.,
  {Szczerba} R.,  {Kidger} M.~R.,    {Lorente} R.,  2012, \apj, 760, 69

\bibitem[\protect\citeauthoryear{{Nenkova}, {Sirocky}, {Nikutta}, {Ivezi{\'c}}
  \& {Elitzur}}{{Nenkova} et~al.}{2008}]{2008ApJ...685..160N}
{Nenkova} M.,  {Sirocky} M.~M.,  {Nikutta} R.,  {Ivezi{\'c}} {\v Z}.,
  {Elitzur} M.,  2008, \apj, 685, 160

\bibitem[\protect\citeauthoryear{{Page} \& {Thorne}}{{Page} \&
  {Thorne}}{1974}]{1974ApJ...191..499P}
{Page} D.~N.,  {Thorne} K.~S.,  1974, \apj, 191, 499

\bibitem[\protect\citeauthoryear{{Petropoulou}}{{Petropoulou}}{2014}]{2014A&A.%
..571A..83P}
{Petropoulou} M.,  2014, \aap, 571, A83

\bibitem[\protect\citeauthoryear{{Potter}}{{Potter}}{2017}]{2017MNRAS.465..337%
P}
{Potter} W.~J.,  2017, \mnras, 465, 337

\bibitem[\protect\citeauthoryear{{Potter} \& {Cotter}}{{Potter} \&
  {Cotter}}{2012}]{2012MNRAS.423..756P}
{Potter} W.~J.,  {Cotter} G.,  2012, \mnras, 423, 756

\bibitem[\protect\citeauthoryear{{Potter} \& {Cotter}}{{Potter} \&
  {Cotter}}{2013a}]{2013MNRAS.429.1189P}
{Potter} W.~J.,  {Cotter} G.,  2013a, \mnras, 429, 1189

\bibitem[\protect\citeauthoryear{{Potter} \& {Cotter}}{{Potter} \&
  {Cotter}}{2013b}]{2013MNRAS.431.1840P}
{Potter} W.~J.,  {Cotter} G.,  2013b, \mnras, 431, 1840

\bibitem[\protect\citeauthoryear{{Potter} \& {Cotter}}{{Potter} \&
  {Cotter}}{2013c}]{2013MNRAS.436..304P}
{Potter} W.~J.,  {Cotter} G.,  2013c, \mnras, 436, 304

\bibitem[\protect\citeauthoryear{{Potter} \& {Cotter}}{{Potter} \&
  {Cotter}}{2015}]{2015MNRAS.453.4070P}
{Potter} W.~J.,  {Cotter} G.,  2015, \mnras, 453, 4070

\bibitem[\protect\citeauthoryear{{Ramakrishnan}, {Hovatta}, {Nieppola},
  {Tornikoski}, {L{\"a}hteenm{\"a}ki} \& {Valtaoja}}{{Ramakrishnan}
  et~al.}{2015}]{2015MNRAS.452.1280R}
{Ramakrishnan} V.,  {Hovatta} T.,  {Nieppola} E.,  {Tornikoski} M.,
  {L{\"a}hteenm{\"a}ki} A.,    {Valtaoja} E.,  2015, \mnras, 452, 1280

\bibitem[\protect\citeauthoryear{{Rani}, {Krichbaum}, {Fuhrmann},
  {B{\"o}ttcher}, {Lott}, {Aller}, {Aller}, {Angelakis}, {Bach} \& et
  al.}{{Rani} et~al.}{2013}]{2013A&A...552A..11R}
{Rani} B.,  {Krichbaum} T.~P.,  {Fuhrmann} L.,  {B{\"o}ttcher} M.,  {Lott} B.,
  {Aller} H.~D.,  {Aller} M.~F.,  {Angelakis} E.,  {Bach} U.,    et al. 2013,
  \aap, 552, A11

\bibitem[\protect\citeauthoryear{{Sironi} \& {Spitkovsky}}{{Sironi} \&
  {Spitkovsky}}{2014}]{2014ApJ...783L..21S}
{Sironi} L.,  {Spitkovsky} A.,  2014, \apjl, 783, L21

\bibitem[\protect\citeauthoryear{{Summerlin} \& {Baring}}{{Summerlin} \&
  {Baring}}{2012}]{2012ApJ...745...63S}
{Summerlin} E.~J.,  {Baring} M.~G.,  2012, \apj, 745, 63

\bibitem[\protect\citeauthoryear{{Tavecchio}, {Becerra-Gonzalez}, {Ghisellini},
  {Stamerra}, {Bonnoli}, {Foschini} \& {Maraschi}}{{Tavecchio}
  et~al.}{2011}]{2011A&A...534A..86T}
{Tavecchio} F.,  {Becerra-Gonzalez} J.,  {Ghisellini} G.,  {Stamerra} A.,
  {Bonnoli} G.,  {Foschini} L.,    {Maraschi} L.,  2011, \aap, 534, A86

\bibitem[\protect\citeauthoryear{{Urry} \& {Padovani}}{{Urry} \&
  {Padovani}}{1995}]{1995PASP..107..803U}
{Urry} C.~M.,  {Padovani} P.,  1995, \pasp, 107, 803

\bibitem[\protect\citeauthoryear{{Zenitani} \& {Hoshino}}{{Zenitani} \&
  {Hoshino}}{2001}]{2001ApJ...562L..63Z}
{Zenitani} S.,  {Hoshino} M.,  2001, \apjl, 562, L63

\bibitem[\protect\citeauthoryear{{Zhang}, {Diltz} \& {B{\"o}ttcher}}{{Zhang}
  et~al.}{2016}]{2016ApJ...829...69Z}
{Zhang} H.,  {Diltz} C.,    {B{\"o}ttcher} M.,  2016, \apj, 829, 69

\end{thebibliography}
\bibdata{References}

\begin{landscape}
\begin{table}
\begin{tabular}{| c | c | c | c | c | c | c | c | c | c | c | c | c | c | c | c | c | c | c |}
\hline
Blazar & $W_{j}$(W) & $L$(m) & $ E_{\m{max}}$(GeV) & $ E_{\m{min}}$(MeV) & $\alpha$  & $\theta_{\m{observe}}$ & $\theta_{\m{opening}}$ & $\gamma_{\m{max}}$ & $\gamma_{\m{min}}$ & $ L_{\m{acc}}$(W) & $ R_{T}$(m) & $ B_{T}$(T)& $x_{T}(\m{pc})$ & $M(M_{\odot})$  \\ \hline 
Markarian 501 & $6.5\times 10^{36}$ & $1\times 10^{20}$ & $228$ & $5.11$ & $1.68$ & $6.5^{0}$ & $5^{0}$ & $9$ & $2$ & n.a. & $1.11\times 10^{14}$ & $8.06\times 10^{-5}$ & 0.304 & $3.16\times 10^{7}$\\ \hline
PKS1510-089 & $4.62\times 10^{38}$ & $2\times 10^{21}$ & $3.22$ & $5.11$ & $1.9$ & $1^{0}$ & $3^{0}$ & $45$ & $35$ & $7.69\times 10^{37}$ & $1.47\times 10^{16}$ & $1.03\times 10^{-6}$ & 40.2 & $4.18\times 10^{9}$\\\hline
PKS1502+106 & $1.94\times 10^{39}$ & $1\times 10^{20}$ & $1.90$ & $5.11$ & $1.8$ & $0.4^{0}$ & $5^{0}$ & $55$ & $38$ & $6.67\times 10^{38}$ & $1.07\times 10^{16}$ & $1.93\times 10^{-6}$ & 29.3 & $3.04\times 10^{9}$ \\ \hline
\end{tabular}
\caption{The values of the physical parameters used for the quiescent fits in the paper, the quiescent fits to Mkn501 and PKS1510 are taken from \citet{2015MNRAS.453.4070P}. Columns from left to right: Fermi blazar name, initial jet power, jet length, electron cutoff energy, injected electron power law index, jet observation angle, jet half-opening angle, bulk Lorentz factor at transition region, bulk Lorentz factor at end of jet, accretion disc luminosity, radius of transition region, magnetic field strength at transition region, distance of transition region and effective black hole mass (if the jet transition region were to occur at $10^{5}r_{s}$ as in M87) also used as the black hole mass for the thin accretion disc fit in FSRQs. \newline
*Please note the corrected typo in the units of $E_{\m{max}}$, which was incorrectly labelled as MeV in \citet{2015MNRAS.453.4070P} }
\label{Table2}
\end{table}
\end{landscape}

\label{lastpage}

\end{document}